\documentclass[8pt,preprint]{aastex}
\def\EGRET{\textit{EGRET}}
\def\FERMI{\textit{Fermi}}
\def\FermiLAT{\textit{Fermi-LAT}}
\def\INTEGRAL{\textit{INTEGRAL}}
\def\INTEGRALSPI{\textit{INTEGRAL/SPI}}
\def\INTEGRALIBIS{\textit{INTEGRAL/IBIS}}
\def\GALPROP{GALPROP}
\def\MUMPS{MUMPS }

\newcommand{\gray}{$\gamma$-ray}

\newcommand{\hi}{H {\sc i}}

\def \phn{ph\ cm$^{-2}$\ s$^{-1}$\ keV$^{-1}$ }
\def \phf{ph\ cm$^{-2}$\ s$^{-1}$ }

\def\diffuse{diffuse\ }
\def\Diffuse{Diffuse\ }

\def\timebins{\textquotedblleft{}time bins\textquotedblright{} }
\def\emptyfield{\textquotedblleft{}empty field\textquotedblright{} }

\def \Ared4.9{\textbf{A$_{4.9\mu}$ }}
\def \factf{($\times 10^{-4}$)}

\shorttitle{Diffuse emission measurement with \INTEGRALSPI}
\shortauthors{Bouchet et al.}

\begin{document}

\title{Diffuse emission measurement with \INTEGRALSPI\ as indirect 
probe of cosmic-ray electrons and positrons}

\author{
  Laurent Bouchet\altaffilmark{1,2}, 
  Andrew W. Strong\altaffilmark{3}, 
  Troy A. Porter\altaffilmark{4},
  Igor V. Moskalenko\altaffilmark{4}, 
  Elisabeth Jourdain\altaffilmark{1,2}, and
  Jean-Pierre Roques\altaffilmark{1,2}
}

\email{bouchet@cesr.fr}

\altaffiltext{1}{Universit\'e de Toulouse, UPS-OMP, IRAP,  Toulouse, France}
\altaffiltext{2}{CNRS, IRAP, 9 Av. colonel Roche, BP 44346, F-31028 Toulouse cedex 4, France}
\altaffiltext{3}{Max-Planck-Institut f\"ur extraterrestrische Physik, Postfach 1603,
85740 Garching, Germany}
\altaffiltext{4}{Hansen Experimental Physics Laboratory, Stanford University,
Stanford, CA 94305}

\begin{abstract}
Significant advances have been made in the understanding of
the  diffuse Galactic hard X-ray  continuum emission using data from the
\INTEGRAL\ observatory.
The diffuse hard power-law component seen with the \INTEGRALSPI\
spectrometer has been identified with inverse-Compton emission from
relativistic (GeV) electrons on the cosmic microwave background and
Galactic interstellar radiation field. 
In the present analysis, SPI data from 2003 to 2009, with a total 
exposure time of $\sim 10^{8}$ s, are
used to derive the Galactic ridge hard X-ray spatial distribution and
spectrum between 20 keV and 2.4 MeV. 
Both are consistent with predictions from the \GALPROP\ code. 
The good agreement between measured and predicted emission from keV to 
GeV energies suggests 
that the correct production mechanisms have 
been identified.
We discuss the potential of the SPI data to provide an indirect probe of the
interstellar cosmic-ray electron distribution, in particular for energies 
below  a few GeV.
\end{abstract}


\section{Introduction}

The Galactic ridge is known to be an intense source of continuum hard 
X-ray and \gray{} emission. 
The hard X-ray emission was discovered by a rocket experiment in 
1972 \citep{Bleach72}, and interstellar emission has subsequently 
been observed from keV to MeV energies by the satellites 
HEAO-1, EXOSAT, Temna, ASCA, Ginga, RXTE,  CGRO/COMPTEL, 
GRANAT/SIGMA, and more recently by Chandra, XMM/Newton, along with INTEGRAL.
Previous analyses of INTEGRAL data showed that, up to 100 keV, a 
large fraction of the total emission from the inner Galaxy is due to 
point sources, and the diffuse emission associated with the 
Galactic ridge is only one tenth of the total Galactic emission in 
the 25--100 keV band 
\citep{Lebrun04, Strong04, Bouchet05, Bouchet08}. 
These analyses also showed that the diffuse emission dominates the 
hard X-ray sky above 300 keV. 

Continuum emission of a diffuse, interstellar nature is expected in 
the hard X-ray and \gray{} regime from several physical processes: 
positron annihilation (through intermediate formation of positronium), 
inverse-Compton (IC) scattering of the interstellar radiation field (ISRF) and 
bremsstrahlung on the interstellar gas 
from cosmic-ray (CR) electrons and positrons, and via 
decay of neutral pions produced in interactions of CR nuclei with the 
interstellar gas. 
Extensive studies of the Galactic diffuse \gray{} 
emission in the context of CR propagation models have 
been carried out by, e.g., \citet{Strong00, Strong04, Strong10} and \citet{Strong2010}. 

In the present analysis, the data accumulated by the SPI spectrometer 
onboard the \INTEGRAL\ observatory are used to derive the spatial morphology
and the spectral shape of the Galactic diffuse emission, 
taking advantage of the greatly increased observational data
and significant advances in the analysis techniques.
This builds on our previous work, described in \citet{Bouchet08} 
and \citet{Porter08}, where we have presented sky maps and 
spectra of the Galactic plane and 
demonstrated the presence of a hard power-law continuum emission, which
was interpreted as IC emission from CR electrons and positrons
upscattering the Galactic ISRF.
For further introductory material and background information on this topic 
we refer the reader to these earlier papers.

\section{Instrument and observations}

The European Space Agency's 
\INTEGRAL\ observatory was launched from Ba\"{\i}konour,
Kazakhstan, on 2002 October 17. 
The SPI spectrometer \citep{Vedrenne03} observes the sky in
the 20~keV to 8~MeV range with an energy resolution ranging from
2 to 8~keV. 
It consists of an array of 19 high-purity germanium
detectors operating at 80~K. 
Its geometric surface area is 508 cm$^{2}$ with a thickness of 7~cm. 
In addition to its spectroscopic capabilities, SPI
can image the sky with a spatial resolution of
$\sim 2.6^\circ$ (FWHM) over a field of view (FoV) of $30^\circ$, because of 
a coded mask located 1.7~m above the detector plane. 
Despite this modest angular resolution, it is possible to locate 
intense sources with an accuracy of a few arc minutes \citep{Dubath05}.
The assembly is surrounded by a 5 cm thick bismuth germanate shield, 
which stops particles arriving from outside the field of view
and measures their flux. 
The instrument's in-flight performance is described in \citet{Roques03}. 
Because of the small number of detectors, SPI's imaging capability relies on a
specific observational strategy, which is based on a dithering
procedure \citep{Jensen03}: the pointing
direction varies around a target in steps of $\sim 2^\circ$ within
a five-by-five square or a seven-point hexagonal pattern. 
In general, a pointing lasts between 30 and 60 minutes, and along its 3~day 
orbit, \INTEGRAL\ operates $\sim$85\% of the time outside Earth's
radiation belts. 
We have analysed observations recorded from Feburary 22, 2003
through January 2, 2009, covering the entire sky.

\subsection{Data selection and preparation}
\label{sec:dataselection}

We exclude data taken during viewing periods (exposures) that
have high background contamination or are dominated
by short transient sources which are not useful for our 
study of the diffuse emission. 
For example, viewing periods containing
solar flares and periods when the spacecraft enters the
radiation belts are excluded from the analysis in order
to remove periods when the data were dominated by high
backgrounds.
For energies around 1~MeV, high-energy particles saturate the electronics 
and can generate false triggers. 
However, it is possible to use these data thanks to other
electronics (via Pulse Shape Discriminators or PSDs) not
affected by the saturation problem. 
These electronics operate in parallel to the fast pre-amplifier output, 
generating an independent trigger for photon energies between
650~keV and 2.2~MeV. 
The trigger signal issued by the PSD electronics is used to confirm and 
select events between 650~keV to 2.2~MeV. 
The procedure is explained in more detail in \citet{JR2009}.

A further selection based on the $\chi^2$ between the sky model and 
the data was made. 
For a few viewing periods 
the sky modelling does not correspond very well with the data. 
This is due to the point-source treatment:

some very short transient sources that are
difficult to identify are missed in our sky model and/or 
there is 
inaccurate modelling of source variability (see Sec.~\ref{sec:pointsource}).

The affected viewing periods, which have relatively high $\chi^2$ values,
contribute only to the systematics for the sky model and hence do not
bring useful 
information on the diffuse emission. 
These viewing periods account for only $\sim$2\% of the whole
dataset.
Therefore, 
we do not use those exposures in the present analysis. 
After their removal, the dataset contains 38699 exposures for the 
analysis below 650~keV,
which corresponds to a livetime of  $\sim1.1 \times 10^8$~s.
For analysis above 650 keV,  
the PSD electronics did not operate during some of
the viewing periods. 
Therefore, we also excluded these
viewing periods, with the total number in our data set for
this energy range then further reduced to 36486. 
The global $\chi^2$ , and the maximum $\chi^2$ per pointing, are given 
in table~\ref{table:exposuresinfolarges} for the best 
model (see Sec.~\ref{sec:galacticridge}).

The data contained in the 196--200~keV and 1336--1342~keV band 
correspond to strong instrumental background lines and are not used in
the analysis.

The first band contains the 198~keV line due to the de-excitation 
of an isomeric state of $^{71}$Ge. 
The second band contains the 1337~keV line related to 
$^{60}$Ge K-shell electron capture, which may blend with the 
1333~keV $^{60}$Fe line. 
We bin the data between 20~keV and 2.4~MeV into 24 energy bands for the 
spectral analysis, and into 7 large energy bands for imaging 
or \diffuse\ continuum morphology analysis. 
For the latter analysis, the data contained in the
1170--1176, 1330--1336 and 1806--1809~keV band are removed because they 
contain counts from the $^{60}$Fe and $^{26}$Al radioactive 
lines (see Sec.~\ref{sec:nonicdiffuse}).

\section{Data analysis}

The signal recorded by the SPI camera on the 
19~Ge detectors is composed of
contributions by each source (point-like or
extended) in the FoV convolved with the instrument response 
(see Appendix~\ref{appendixA}), 
plus the background.
For $N_s$ sources located in the FoV, the counts 
$D_{p} ^d$ obtained during an
exposure (pointing) $p$ in detector $d$ for a given energy band, can be 
expressed by the relation

\begin{equation}
D_{p} ^d=\sum_{j=1}^{N_s} R_{p,j} ^d S_{p,j} + B_{p} ^d
\label{eq1}
\end{equation}
   
\noindent 
where $R_{p,j} ^d$ is the response of the instrument for the source $j$, 
$S_{p,j}$ is the flux of the source $j$, and $B_{p} ^d$ is the background 
recorded of the pointing $p$ for detector $d$. 
For a given pointing $p$,  $D_{p} ^d$, $R_{p,j} ^d$, 
and $B_{p} ^d$ are vectors of $N^d$, here 19, elements.  
For a given set of $N_p$ exposures, we have to solve a system of
$N_p \times N ^d$ equations (Eq.~\ref{eq1}). 
For extended/diffuse sources, we assume a spatial morphology given 
by an analytic function or a model sky map. 
To reduce the number of free parameters related to 
the background, 
we take advantage of the  stability of relative count rates between detectors
to rewrite the background term as

\begin{equation}
B_{p} ^d = A_p \times U^d \times t_{p} ^d
\label{eq2}
\end{equation}

\noindent
where $A_p$ is a normalisation coefficient per pointing related to the 
background intensity, $U^d$ represents the background count rate pattern
on the SPI camera for the $d^{th}$ detector ($U^d$ is a 19-element vector),
and 
$t_{p} ^d$ is the effective observation time for pointing $p$ and detector $d$. 
The number of parameters necessary to model the background reduces 
to $N_p$ [$U$ is assumed to be known and can be, in some cases, 
determined independently from \emptyfield 
exposures (see Sec.~\ref{sec:background})].

The number of free parameters in the set of $N_p \times N^d$ equations is 
then $N_p \times N_s~ +~ N_p$ (for the $N_s$ sources and
the background intensities).
An additional reduction of the number of parameters 
can be obtained because many sources vary on
timescales larger than the exposure timescale. 
Furthermore, many point sources are weak enough to be considered
as having a constant flux within the statistical errors, especially for 
higher energies.
Then the $N_p \times N_s$ parameters related to the sources are 
reduced to $N_s^{eff}$ parameters. 
For example,
in the 25--50~keV range, for 257~sources and 38699~pointings, 
$N_s^{eff} \simeq$  22500 \timebins.

\subsection{Background}\label{sec:background}

The modelling of the instrumental background is an important issue and 
challenge for the data analysis. 
The distribution of the instrumental
background in the detector plane changes significantly during the
period spanned by the observations because 2 detectors failed early in the 
mission (detector 2 on December 7, 2003 and detector 17 on July 17, 2004). 
The ``uniformity''
map or background pattern (Eq.~\ref{eq2}) should ideally be derived from
\emptyfield observations. 
However, the
dedicated \INTEGRALSPI\ \textquotedblleft{} empty fields\textquotedblright{}
are too rare to
derive suitable and precise ``uniformity'' maps for the large dataset
used in the present analysis.
Another way to determine the background pattern is to solve Eq.~\ref{eq1} for
the intensities of both sources, background, and detector pattern
simultaneously.
 
The resulting patterns are therefore model dependent because of the
unavoidable cross-talk between the extended source emission and the 
background pattern (and sources).
These effects have been intensively studied above and around 500~keV,
where many high latitude exposures can be considered as
\emptyfield exposures \citep[see][]{Bouchet10} 
because they contained no detectable sources. 

For the present study this has a minor impact because the 
sky model is sufficiently well approximated and hence the derived intensities
are not significantly affected.

We treat the background intensity as varying with time
and test several timescales. 
Timescales above 6 hours give poor $\chi^2$.
In general, the F-test shows that using a background 
intensity, $A_p$ (Eq.~\ref{eq2}), which varies on an exposure 
timescale ($\sim$2800 s), 2 or 4 hours does not improve the fit compared
to using a background timescale of $\sim$6 hours.
The $\sim$6~hours timescale analysis also produces slightly 
reduced error bars because there are less free parameters 
for background and source components.
However, the
intensities are perfectly compatible with those obtained with a more
variable background (timescale $<$ 6 hours).
We therefore use a background timescale of $\sim 6$ hours throughout the 
following analysis.

Finally, the standard configuration used for the  analysis
consists of computing the background pattern per period between two
detector annealings (performed every $\sim$6~months to restore the detector energy resolution)
 with an intensity normalisation 
that varies with a 6-hour timescale
(leading to $\sim$6000 intensities to be fitted for the 38699 exposures).

\subsection{Sky modelling}

The data contains contributions from point sources, positron annihilation
radiation (511 keV line plus positronium continuum) and from
the process of interest in this study, the ``diffuse'' (non-annihilation) 
continuum emission, in addition to backgrounds.
To extract the \diffuse flux with SPI, information for the source positions 
along with their variability timescales is required. 
We therefore need to build a sky model to derive the corresponding fluxes.

\subsubsection{Point-source emission}
\label{sec:pointsource}

We follow a similar method to that described in \citet{Bouchet08}
where the point-source contributions are simultaneously extracted together
with the \diffuse and background components.
For
each source, the algorithm divides the total observation time into
smaller 
time intervals, down to the exposure duration, where the source can be considered
to have a constant flux. 
The algorithm, based on signal segmentation
\citep{Scargle98}, can detect localised time
structures and generally characterise intensity
variations. 
It allows the construction of \timebins intervals for each
source and energy band. 
The {\it a priori} information can thus be
introduced in the source terms through their position and variability
timescales.
The intensities in these \timebins are parameters to be fitted. 
The catalogues
derived in this way optimise the signal-to-noise ratio of both
sources and diffuse emission.

However, it is difficult to take into account all transient sources,
especially on the shortest timescales
\citep{Bird10}. 
The information about
precise burst locations in time were not available at the time of this
analysis and/or is difficult to deduce from the data directly in a blind 
search.
The consequence of missing these sources is to reduce slightly the
number of exposures used here, because exposures that are not 
well fit are excluded from the final dataset 
(see Sec.~\ref{sec:dataselection}). 

\begin{figure}
\plotone{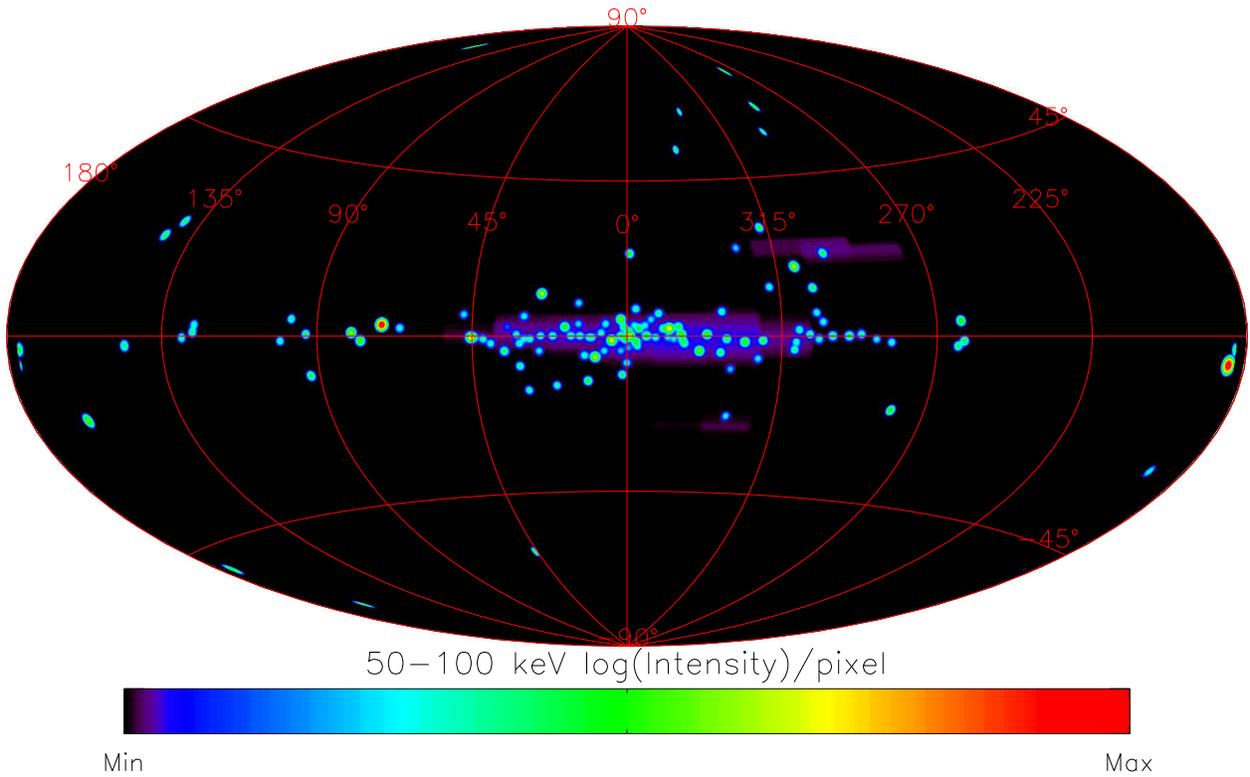}
\caption{50--100~keV intensity sky map including both sources and diffuse 
emission. 
The diffuse part of the image has pixels of different size and is first 
downsampled to a common pixel size of $3^{\circ} \times 2.6^{\circ}$, 
thereafter the image is smoothed by a $3\times3$ pixels boxcar.}
\label{fig:allsky_image50_100}
\end{figure}

The resulting \INTEGRALSPI\ all-sky survey allows us to identify 254
point sources in the 25--50~keV energy band, with 123, 53, and 26 of the
point sources identified at in the lower energy band still emitting 
in the 50--100, 100--200, and 200--600~keV bands, respectively.
See Tables~\ref{table:exposuresinfolarges} 
and~\ref{table:exposuresinfospectral}. 
Figure \ref{fig:allsky_image50_100}
shows the 50--100~keV all sky map.

For the spectral analysis, a
given source is considered to emit as long as its spectrum is detected
above $1\sigma$. 
Table \ref{table:exposuresinfospectral} gives the number of point sources 
used to model the data in the different energy bands. 
More details on the point-source analysis will be given in
a forthcoming paper, while in the present work we
focus on the diffuse emission.

\subsubsection{Annihilation radiation spectrum}
\label{sec:annihilation}

Between 250~keV and 511~keV annihilation radiation is the dominant
emission process with two components associated with the Galactic bulge and
disk.
We assume that the spatial morphologies for the Galactic bulge 
annihilation line and positronium continuum are the same. 
The bulge morphology 
is described by a combination of $3.2^\circ$ and $11.8^\circ$ 
Gaussians centred at $l = -0.6^\circ$ and $b = 0.0^\circ$ with the ratio of the
$11.8^\circ$ to the $3.2^\circ$ Gaussian flux fixed to 2.16 \citep{Bouchet10}. 
For the disk, we follow the \citet{Bouchet10} modelling for the 511~keV 
line emission using either 
a 240 $\mu$m NIR/DIRBE\footnote{http://lambda.gsfc.nasa.gov} map
or a halo morphology \citep[see also][]{Weidenspointner08,Churazov10}.
Note that in the present analysis we are performing a similar analysis to
that described in \citet{Bouchet10}, but over a wider energy range for the 
annihilation radiation line and ortho-positronium continuum emission.

The bulge spatial morphology is sufficiently
distinct from the \diffuse 
continuum (plus annihilation disk) spatial morphology to be
extracted separately from the data,
without suffering from ``cross-talk''.
On the other hand, the disk annihilation continuum is weak compared to
the ridge emission and too close in spatial shape to be distinguishable. 
Consequently, the disk contribution cannot be directly measured, but only 
inferred from the \diffuse continuum measurements (Sec.~\ref{sec:nonicdiffuse}).

\subsubsection{\textquotedblleft{}Diffuse\textquotedblright{} emission}

To derive the spatial morphology of
the \diffuse emission (diffuse continuum and annihilation radiation), 
we use two approaches.
The first is sky ``imaging'' and the second involves sky model fitting.

For ``imaging'', sky pixels containing the flux of the \diffuse emission
(sum of annihilation radiation plus diffuse continuum) are built. 
The point source locations
are then introduced separately as {\it a priori} 
information \citep{Bouchet08}. 
Intensities of both diffuse
pixels and point sources are parameters to be fitted.

For sky model fitting, several
parameterised model distributions representing the annihilation
radiation spectrum and diffuse continuum emission are directly tested 
against the data. 
The spatial morphology of the annihilation radiation spectrum, diffuse 
continuum emission, and point source locations  are
introduced separately as {\it a priori} information.  
Intensities and fluxes for all the components are parameters to be fitted.

In both cases, sky parameters and background intensities, along with detector 
patterns, are fitted to
the data through the response matrix to maximise the likelihood
function.

\subsection{Core algorithm}
\label{sec:corealgorithm}

The tools used for background modelling, imaging, and model fitting were
specifically developed 
for the analysis of SPI data and described in \citet{Bouchet08}.
To determine the sky model parameters we adjust the data through a 
multi-component
fitting algorithm, based on 
the likelihood test statistic. 
The core algorithm to
handle such a large, but sparse, system is based on \MUMPS\footnote{\MUMPS 
(MUltifrontal Massively Paralell Solver) software developed by the \textit{IRIT/ENSEEIHT} laboratory
(http://mumps.enseeiht.fr)} \citep{Amestoy06} together with an error 
computation
method
dedicated and optimised for the \INTEGRALSPI
response matrix structure \citep{Rouet09, Tzvetomila09}.

\section{\textquotedblleft{}Diffuse\textquotedblright{}~continuum analysis results}

\subsection{Galactic Emission Spatial Morphology}
\label{sec:ridgespatial}
\begin{figure}
\plotone{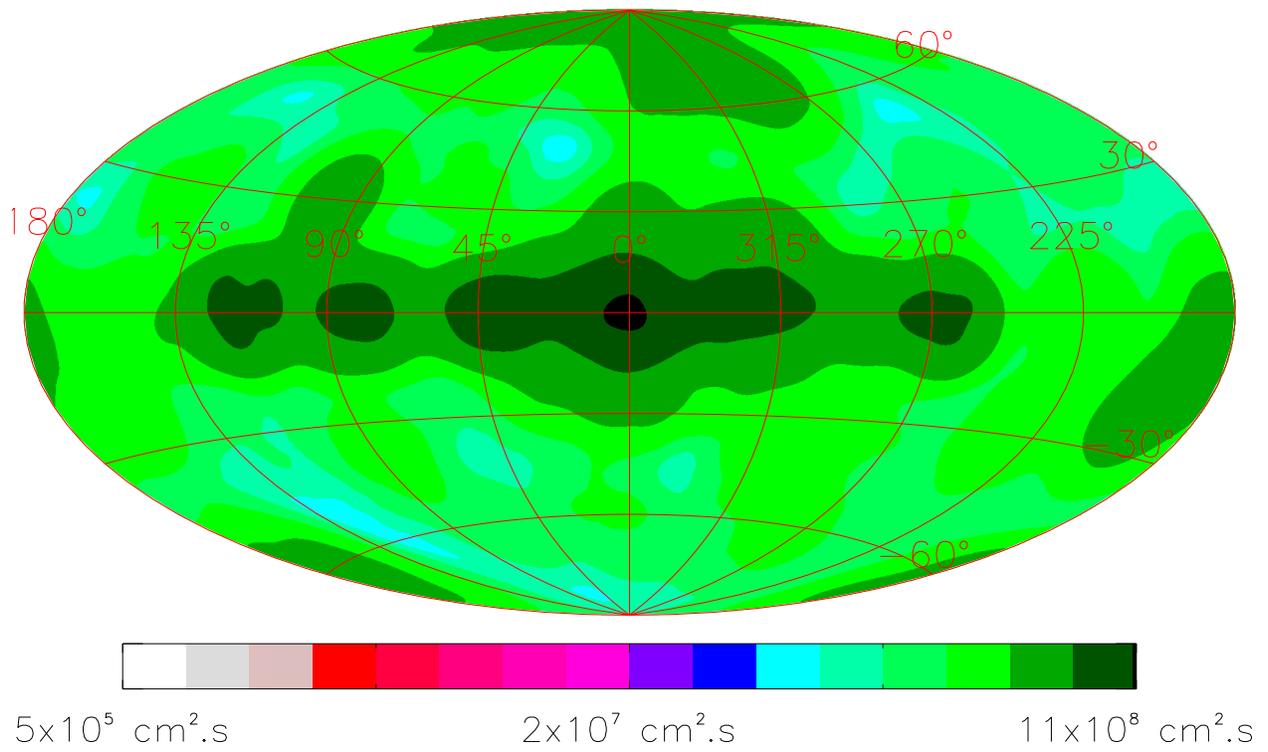}
\caption{25--600~keV \INTEGRALSPI~exposure map in cm$^2$ s. 
This map takes into account the
differential sensitivity of SPI across its field of view.}
\label{fig:exposition_map}
\end{figure}

\begin{figure}
\plotone{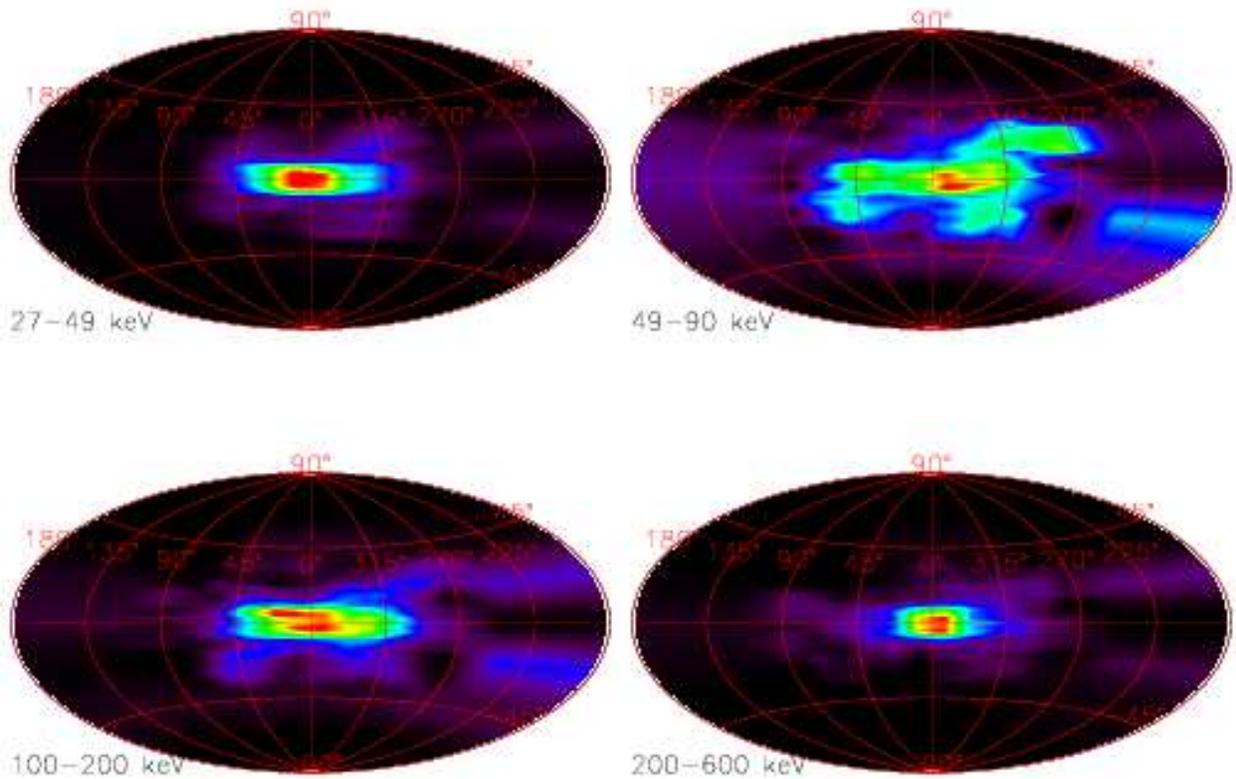}
\caption{\textquotedblleft{}Diffuse\textquotedblright{} emission 
intensity all sky maps. 
The energy bands are
(top-left) 27--49~keV, (top-right) 49--90~keV,
(bottom-left) 100--200~keV and (bottom-right) 200--600~keV.
The original images have pixels of different size and are first downsampled 
to a common pixel size of $3^{\circ} \times 2.6^{\circ}$, 
thereafter they are smoothed by a $3\times3$ pixels boxcar. 
The \textquotedblleft{}Diffuse\textquotedblright{} emission 
energy is minimum in the 50--100~keV band. 
The color intensity is the same as in Fig.~\ref{fig:allsky_image50_100}.}
\label{fig:allsky_image_diffuse}
\end{figure}

To determine the (model independent) spatial distribution of the
Galactic Ridge emission, the entire sky is divided into pixels. 
The sky
exposure (Fig.~\ref{fig:exposition_map}) is extremely non-uniform. 
To compensate, the pixels
are chosen to have different sizes in latitude and longitude
to extract fluxes with a comparable signal-to-noise ratio in each pixel
over the whole sky. 
Because the point sources intensities are derived simultaneously,
the pixel sizes also have to be chosen large enough to avoid ``cross-talk'' 
with point sources.
The region
$\vert l \vert < 100^{\circ}$,
$\vert b \vert < 30^{\circ}$
is divided into cells of size 
$\delta l \times \delta b = 15^{\circ} \times  2.6^{\circ}$ 
($ 40^{\circ} \times  5.5^{\circ}$ for the 600--1800~keV 
band).
Outside of this region, the pixels have a larger sizes depending on 
their locations.
The pixel sizes are chosen {\it a posteriori} to optimise
the signal-to-noise ratio per cell while being sufficiently small to
follow the observed \diffuse spatial variations. 
The number of unknowns (the diffuse part of the sky to be 
\textquotedblleft{}imaged\textquotedblright{}
contains 201 pixels for $E < 600$~keV and 121 pixels above)
is high but reasonable compared to the available
data, and the problem is easily tractable using a simple likelihood
maximisation
to determine all the corresponding intensities and error estimates. 
For this particular analysis, the fitted intensities are constrained to be 
positive.
This constraint stabilises the ``imaging'' system of equations to be solved.
Figure \ref{fig:allsky_image_diffuse}
shows the all-sky intensity images of the \diffuse emission in several
energy bands.
These images and profiles give a qualitative view of the \diffuse
emission, but very localised structures outside the region delimited by
$\vert l \vert > 100^{\circ}$ and
$\vert b \vert > 30^{\circ}$
may be missed, due to the sizes of the pixel in this region
$\delta l \ge 15^{\circ} \times \delta b \ge 5.2^{\circ}$. 

\begin{figure}
\epsscale{0.7}
\plotone{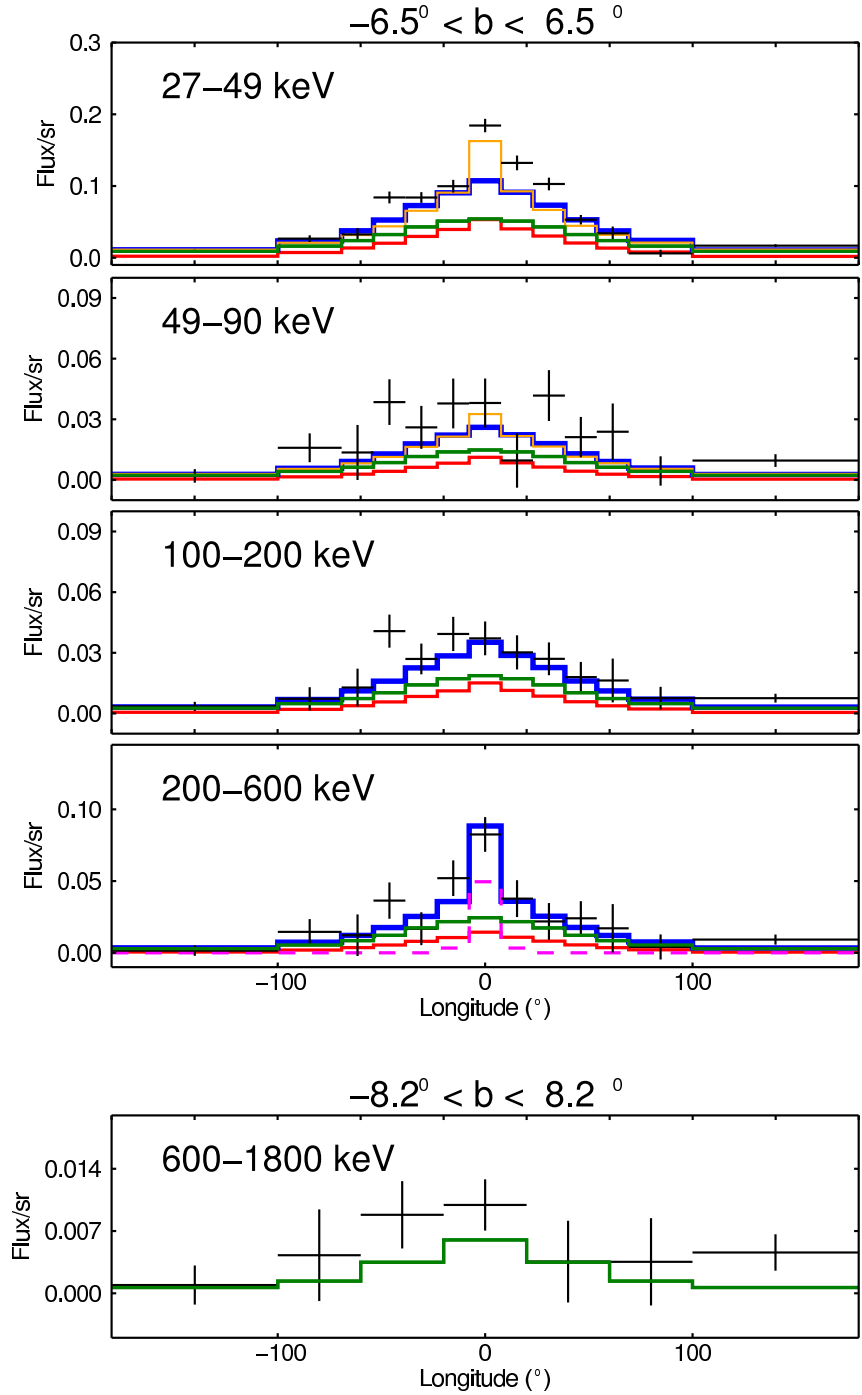}
\caption{Longitude profiles of the Galactic Ridge emission in different
energy bands, observed with \INTEGRALSPI (crosses). The best fit template maps 
consiting of a combination of \GALPROP\ IC distribution,  
A$_{4.9 \mu {\rm m}}$, and the annihilation radiation 
contribution (table~\ref{table:diffusebestmaps}) is shown in blue. 
A$_{4.9 \mu {\rm m}}$ (stellar emission) 
and IC contributions are shown respectively in red and green.
The pink dash-dotted line is the annihilation radiation contribution in the 
200--600~keV energy band.
The orange thin line (best absolute model) is a model consisting of IC 
plus DIRBE-based 4.9 $\mu$m template for energy 
below 49~keV and IC plus 60$\mu$m emission map between 49~keV and 200~keV.
}
\label{fig:profile_longitude}
\end{figure}

\begin{figure}
\plotone{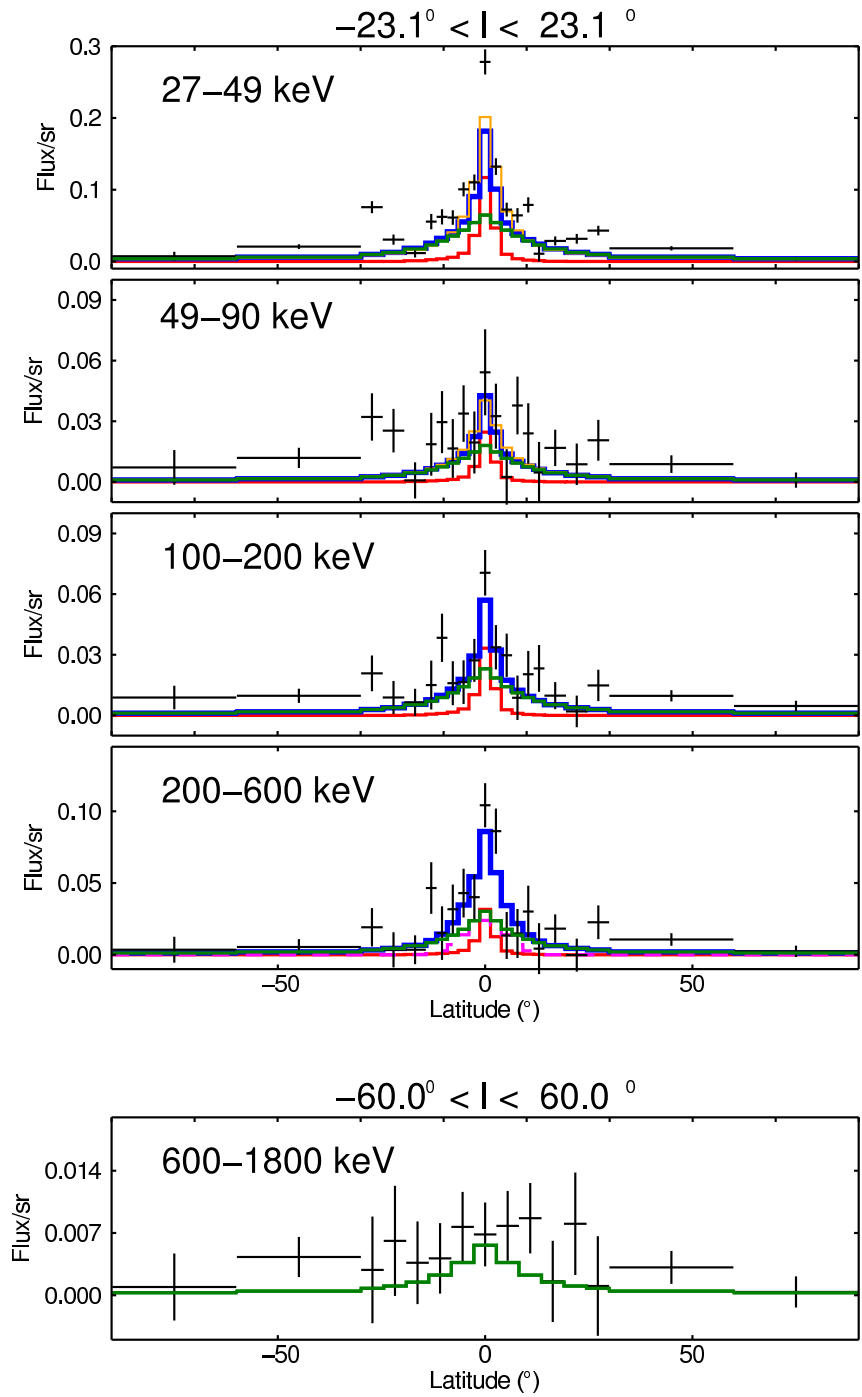}
\caption{Same as Fig.~\ref{fig:profile_longitude}, but for Galactic latitude.}
\label{fig:profile_latitude}
\end{figure}

To quantify more easily the behaviour of the \diffuse emission, we
present our results in terms of longitude and latitude profiles in 
Fig.~\ref{fig:profile_longitude} and~\ref{fig:profile_latitude}.
These figures are obtained by integrating the flux measured for
$\vert b \vert \le 6.5^{\circ}$ in longitude 
or $\vert l \vert \le 23^{\circ}$ in latitude, and  
$\vert b \vert \le 8.2^{\circ}$ and $\vert l \vert \le 60^{\circ}$ for 
the 600--1800~keV band.

Because the images and profiles have a low signal to noise ratio, in order
to have more quantitative measurements, a complementary analysis was 
performed in which the \diffuse
continuum spatial morphology is modelled using several template 
maps (table~\ref{table:diffusebestmaps}), separately or in combination.

Several DIRBE emission maps have been tested as a tracer of 
the \diffuse emission: the NIR/DIRBE (1--10 $\mu$m) related to the stellar 
emission, the MIR/DIRBE
map (10--30 $\mu$m) related to dust nanograins and PAHs heated to high 
temperatures, and the 
FIR/DIRBE (30--240 $\mu$m) related to $\sim$ micron sized dust grains emitting
in thermal equilibrium with the heating Galactic ISRF.
We have also tested 21-cm \hi\ \citep{Dickey90} and $^{12}$CO \citep{Dame01} 
maps.
The
CO template traces the 
molecular gas of the Galaxy (which can be related to the distribution to
the young stellar populations), and are 
used to represent the spatial distribution for the nucleosynthesis lines. 

Our analysis employs extinction-corrected IR maps, where the method is  
described in \citet{Krivonos07} which produces corrected maps to an 
accuracy of $\sim10$\%.
We denote these maps as A$_{\lambda}$ where $\lambda$ is the wavelength.
For example, the A$_{4.9 \mu{\rm m}}$
map is motivated by the probable stellar origin for the Galactic ridge emission
below 100 keV \citep{Krivonos07}. 

The IC maps are based on a physical model for the IC emission generated with 
the \GALPROP\ code (Sec.~\ref{sec:galprop}).
Note that the IC map spatial morphology is energy dependent, whereas 
the spatial morphology for the other templates is independent of energy band.

All template skymaps are convolved with the instrument response 
and compared to the data. 
The resulting fit parameters are given in table~\ref{table:diffusebestmaps}.

If a single template is fitted (with intensity as free parameter), the 
IC templates (models 54z04LMS  or 54z04LMS-efactor, described in 
Sec.~\ref{sec:ICspec})  
are the best tracer for the spatial distribution of the 
diffuse continuum.
They provide a better fit to the whole-sky distribution because they 
account for the emission at high latitudes, but are not always the best fit 
in the ridge region, where the NIR/DIRBE  maps fit the emission better. 
The best non-IC map is the A$_{4.9 \mu{\rm m}}$, 
especially for the low energy range(E $<$ 200~keV), consistent with 
the results of \cite{Krivonos07} who found that this map traces the \diffuse 
stellar emission distribution at low energies.
The IC templates constitute the best model of the  whole sky
``diffuse'' emission, but our derived map contains some
unresolved source component in the ridge region. 
Therefore, an additional second map is needed to model
more accurately the excess in the region $\vert b \vert < 10^{\circ}$.

Above 600 keV, the IC map fits the data very well, although a marginally 
better fit can be obtained with the 1.25~$\mu$m map.
However, the improvement over the IC map is small and is restricted to this
energy range.
A combination of
IC and 1.25~$\mu$m maps is also marginally 
a better fit to the data in the 600--1800~keV band. 
But, as similarly for the other possible combinations of maps,
does not really improve the fit, considering the number of extra degrees 
of freedom introduced. 
Furthermore, when the maps are combined the contribution of each separate 
map is difficult to measure, 
due to both the statistics and/or the unavoidable 
\textquotedblleft{}cross-talk\textquotedblright{}. 
So, above 600 keV, qualitatively, the best \diffuse spatial emission model 
is the IC map. 

At energies below 600 keV, the situation is more complex. 
In addition to the diffuse continuum, the annihilation radiation component
is modeled as described in Sec.~\ref{sec:annihilation}.
A single IC map is not sufficient to fit the rest of the emission 
distribution, because it
does not fit well the peak along the Galactic plane.
An additional map such as DIRBE, CO, or HI is needed to reproduce this 
structure.  
Table~\ref{table:diffusebestmaps} indicates that below 600 keV, better fits 
to the spatial morphology emission can be obtained by
combining two maps. 
For the 25--600~keV band, a combination of IC and the 4.9 $\mu$m, 
3.5 $\mu$m or A$_{4.9 \mu{\rm m}}$ map gives the best fits.
For the 25--50~keV band an overall best fit can be obtained with a 
combination of IC and the 4.9 $\mu$m maps.
This is consistent with the stellar origin for the Galactic ridge 
emission in the X-ray domain proposed by \citet{Krivonos07}.

Based on these results, and to simplify the statistical analysis, the 
emission profiles are finally fitted with a combination of 
the A$_{4.9 \mu{\rm m}}$ plus IC and bulge 
maps below 600~keV, and with a pure IC map above 600~keV, see 
Figs.~\ref{fig:profile_longitude} and~\ref{fig:profile_latitude}, and 
table \ref{table:diffusebestmaps}. 
With this method 
the signal-to-noise ratio of the derived \diffuse total emission is optimised.
Using the absolute best-fit model for each narrow energy band (bold-italic 
in table~\ref{table:diffusebestmaps})
instead of the above adopted model (bold in 
table~\ref{table:diffusebestmaps}), does not change
the results significantly.

\subsection{Galactic Ridge Spectral Analysis}
\label{sec:galacticridge}

In the previous section we considered the spatial morphology and choice 
of skymap templates.
We now turn to the spectral analysis.
To extract spectra, we fix the sky model so 
that the diffuse spectrum morphology below 732 keV 
(the relevant upper spectral channel above 600 keV), is modelled with a linear 
combination of A$_{4.9 \mu{\rm m}}$ and an IC map computed using \GALPROP, 
plus two Gaussians representing the bulge annihilation radiation component. 
For energies above 732~keV, the morphology is modelled by the IC map 
plus 3 spectral lines (Sec.~\ref{sec:nonicdiffuse}).
The intensities of the spatial components related to the diffuse 
emission together with sources and background intensities are adjusted 
to the data as described in Sec.~\ref{sec:corealgorithm}.

Details on the individual component separation procedure are given in 
Appendix~\ref{appendixA}.
The bulge morphology differs significantly from the A$_{4.9 \mu {\rm m}}$ 
and IC map components and there is very little 
``cross-talk'', hence this component is easily extracted
separately (Table~\ref{table:diffusebestfits}a). 
On the other hand, IC and A$_{4.9 \mu {\rm m}}$ components are 
more difficult to disentangle due to their rather similar morphologies.
Nevertheless, it is possible to separate the contribution of each map 
(Table~\ref{table:diffusebestfits}b-c).
To take into account the ``cross-talk'' between the IC and A$_{4.9 \mu {\rm m}}$ 
components, we add, in addition to statistical errors on fluxes, 
artificial systematic errors\footnote{ 
The reduced $\chi^2$ between the data and the model is abnormally high ($\chi^2  > 2$) 
and technically XSPEC (v11) spectral fitting code does not compute the standard 
deviation of the fitted parameters. "Cross-talk" produces data fluctuations 
that are not statistical; systematic errors are then added to statistical errors. 
Adding 25 $\%$ systematic error in XSPEC gives a reduced $\chi^2$ of about 1. These systematic mainly increase errors bars below 100 keV.
}.
When deriving these components, 
the data above $\sim$ 1~MeV are not used because the lower statistics do not
allow for the distinction.
The IC and A$_{4.9 \mu {\rm m}}$ spatially-derived maps 
are then fitted as a superposition of three expected physical 
spectral components:

\begin{itemize}
\item Emission by ``unresolved sources'', dominating below $\sim 50$~keV, 
modelled by 
an exponential cutoff power-law spectrum.

\item Annihilation radiation spectrum modelled using a 
Gaussian centered at 511~keV with 2.5~keV 
FWHM plus positronium continuum \citep{Ore49}.

\item Diffuse continuum mainly attributed to interstellar emission, modelled 
by a power law.

\end{itemize}

The diffuse spectral fitted parameters following the above procedure 
for the central radian defined 
by $\vert l \vert < 30^{\circ}$, $\vert b \vert < 15^{\circ}$ are given in 
table \ref{table:diffusebestfits}.

The IC extracted component is found to have a
power-law index $\sim$ 1.8 with a 100~keV flux $9\times 10^{-5}$ \phn.
The A$_{4.9 \mu {\rm m}}$ tracer component 
is found to have a much harder power-law index of $\sim$ 1 with an intensity 
significantly lower than the IC component.  
In other words, the IC component dominates the flux over the sky.
For the full sky, the ratio of IC to A$_{4.9 \mu {\rm m}}$ components   
is 14, 8, 1.8 and 0.95 at 50, 100, 500 and 1~MeV, respectively. 
When considering only the central radian,
the IC 
component contains 15$\%$ at 20~keV rising 
to 25\% at 1~MeV of the total IC 
sky flux. 
The 20--100~keV flux ratio of IC to A$_{4.9 \mu {\rm m}}$  component 
is $\sim$9 for the whole sky and 2.6 for the 
central radian. 

\begin{figure}
\plotone{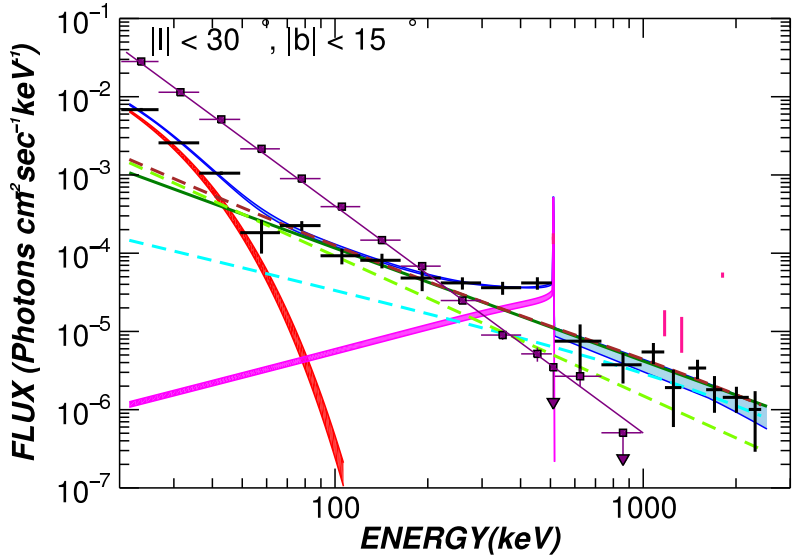}
\caption{Spectra of the different emission components in the central
radian of the Galaxy
($\vert l \vert < 30^{\circ}$ and $\vert b \vert < 15^{\circ}$). 
SPI measurements are black crosses.
Violet squares: total emission of resolved sources. 
Violet line: power law fit to the
resolved sources emission (power-law index of 2.9 and flux at 
100 keV $4 \times 10^{-4}$ \phn).
Blue: total diffuse emission -- Magenta:
annihilation radiation spectrum (line + positronium) \textendash{} Red:
Emission of low energy \textquotedblleft{}unresolved\textquotedblright{} 
sources.
The possible range of variation of these components are represented with the 
shaded area.
Dark green line \textendash{} is the deduced
continuum emission thought to be dominated by CRs interacting in the ISM.
The diffuse continuum best fit spectrum  based both on spatial morphology 
and spectral decomposition
is indicated by the dashed cyan (A$_{4.9 \mu {\rm m}}$ spatial component) 
and green dashed lines (IC component). 
The sum of these two
components is the brown dashed line which compared to the power-law fit with 
index 1.44 based solely on spectral decomposition (dark green line).
}
\label{fig:diffuse_spectrum}
\end{figure}

The final fit is made for the total spectrum (sum of all components) 
by combining the spectral information obtained from each spatial morphology 
component fit.
For this step,  
the A$_{4.9 \mu {\rm m}}$ and IC power-law indices and intensities are fixed 
while fitting the other components. 
We also introduce a cutoff in the A$_{4.9 \mu {\rm m}}$ component to steepen 
its contribution at higher energies (above 1~MeV) where it would be 
inconsistent with the data from the higher energy instruments otherwise.
We model the component related to
unresolved sources with a single exponential cutoff spectrum. 
If the summed spectral components for IC, A$_{4.9 \mu {\rm m}}$, and annihilation
radiation are directly decomposed into the three spectral components 
described above, then the resulting spectral shape is similar to that described in the previous paragraph. 
The interstellar emission component is then best fit by a power-law of 
index $\sim$ 1.4-1.5.
It results in a diffuse continuum spectrum very similar to that of the summed power laws of the IC and A$_{4.9 \mu {\rm m}}$ components. 
The total \diffuse spectrum 
for the central radian region is
presented in Fig.~\ref{fig:diffuse_spectrum}. 

These results are consistent with our previous work \citep{Bouchet08}, 
except that the diffuse continuum 
power law obtained with the IC maps has a higher intensity 
(20--30$\%$ at 50~keV). 
This is due to the large latitude extent of this model which 
contributes significant integrated flux very far from the Galactic 
ridge, and not included in \citet{Bouchet08}. 

Above 1~MeV, the decomposition of the emission 
into two spatial components gives too much uncertainty for each component.
We therefore show our determination of the diffuse emission based on estimating
a minimum and maximum extracted intensity for each component.
To do this we fitted the spatial morphology with the A$_{4.9 \mu {\rm m}}$ tracer,
which  gives  nearly the minimum extracted flux, 
while the IC map gives the maximum extracted flux.
A similar spectral analysis as that described above was then performed. 
The range of uncertainty is shown as a shaded area 
in Fig.~\ref{fig:diffuse_spectrum}.

\subsection{\Diffuse emission contamination with extragalactic background emission}

The conventional coded-mask system provides, by construction, flux free 
from extragalactic backgound (EGB) contamination, but for INTEGRAL/SPI we 
need a background model (see Sec.~\ref{sec:background}).
Due to the background modelling there is uncertainty
associated with the level of diffuse extragalactic background (EGB) flux in 
our diffuse emission determination.
The instrument background is modelled with a background pattern whose 
intensity varies on a $\sim$ 6 hours time scale (Sec.~\ref{sec:background}).
Alternatively, it can also be modelled as above but with  an additional 
isotropic term, which can be either stable in time (to represent the EGB) 
or variable in time 
(related to long term variations of the background component).
Nevertheless, it is difficult to distinguish between the two, and the 
first one is preferred for its simplicity. 

In our analysis, for the region $\vert l \vert \leq 30^\circ$ 
and $\vert b \vert \leq 15^\circ$ we find that EGB contamination is 
negligible ($< 3\%$)
below 600~keV.
For energies above 600~keV, because of the lower statistics, 
if we assume that the measured diffuse emission is EGB-dominated, then 
the diffuse component is reduced by at most  $\sim 28 \% $
in the region $\vert l \vert < 30^\circ$ and $\vert b \vert < 15^\circ$.
This is within the error bars and does not affect our conclusions associated
with the modelling (see below).

\section{Modelling the \diffuse emission}
\subsection{\GALPROP\ models}
\label{sec:galprop}

The \GALPROP\ code \citep{Strong98,Strong00, Strong04, Strong07} 
including a new model for the Galactic
interstellar radiation field (ISRF) is the basis for predicting 
Galactic \diffuse emission in the energy range from keV to TeV energies,
thus covering more than 10 orders of magnitude in energy
\citep{Porter08}. 
The goal of the \GALPROP\ project is to 
combine CR and broadband diffuse emission data from radio to \gray{s} into a
single interpretative framework.
Therefore, while the modelling and interpretation has most recently 
focussed on \gray{}
data from the \FERMI\ mission, it is also applicable
to other experiments like {\it WMAP}, {\it PLANCK}, \INTEGRAL\ 
and {\it MILAGRO}.

In \citet{Porter08} we used the so-called ``EGRET-excess'' based model \citep{Strong04}, 
which invoked CR proton and electron spectra different from those
measured directly in the Solar System, in order to account for the \gray{} 
spectrum measured by \EGRET. 
Now that the ``GeV excess'' in this spectrum has been
shown to be absent in \FermiLAT\ data \citep{Abdo09}, being presumably 
an \EGRET\ instrumental effect, we use in the present work
the ``conventional model'', which requires consistency of the modelled CR 
intensities and spectra with those directly measured.
We use the model (GALPROP ID 54\_z04LMS) described in \citet{Strong10}, 
which reproduces the electron (plus positron) spectrum measured by \FermiLAT\
\citep{Abdo10}, but is not fitted to \FermiLAT\ \gray{} data. 
It has a halo height of 4 kpc and includes CR reacceleration; for further 
details see \citet{Strong10}.

\begin{figure}
\plotone{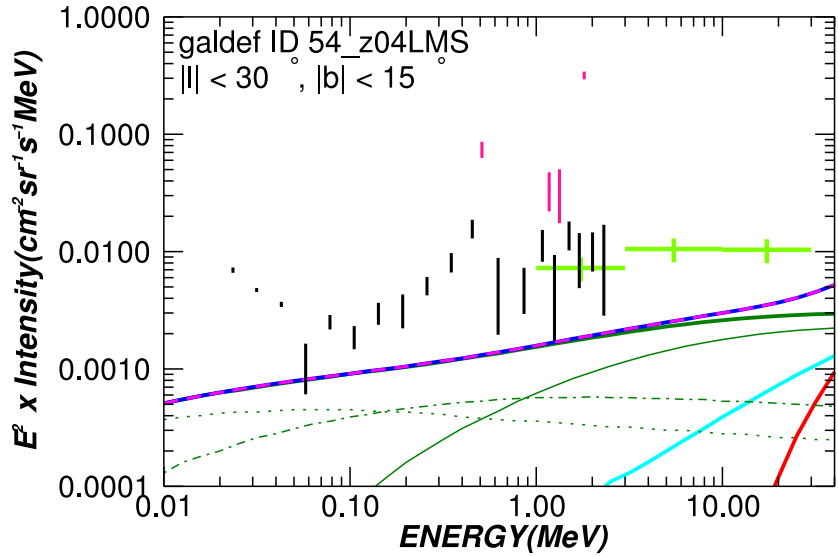}
\caption{Spectrum of the diffuse emission for
($\vert l \vert < 30^{\circ}$ and $\vert b \vert < 15^{\circ}$).
Black data points are for SPI
(positron annihilation, $^{26}$Al and $^{60}$Fe lines are shown in red-pink).
Green data points are for
COMPTEL.
The blue/pink line is the total emission as calculated with the \GALPROP\ 
code, with the primary electron spectrum based on \FermiLAT\ measurements.
Green solid line: total Inverse-Compton emission (IC) --
Red solid line : $\pi^0$-decay -- green thin line : IC (optical) -- Green
short-dash thin line IC (IR) -- Green dotted thin line: IC (CMB) -- Cyan 
solid line : bremsstrahlung emission. }
\label{fig:galdef_id54_z04LMS}
\end{figure}

The calculations presented in \citet{Strong04}, \citet{Porter08} and \citet{Strong10} 
show the importance of secondary leptons in CRs for
the proper calculation
of the \diffuse emission. 
Secondary CR positrons and electrons produced via
interactions of energetic nucleons with interstellar gas are usually
considered as a minor CR component.
However, the secondary positron and electron flux is comparable to the 
primary electron flux around $\sim 1$~GeV in the ISM, providing diffuse 
emission
in addition to that from primary CR electrons.
The enhancement is $\sim 1.2-1.4$ times higher in the IC
\gray{s} up to MeV energies relative to that from
pure primary electrons.  
This
leads to a considerable contribution of secondary positrons and
electrons to the diffuse \gray{} flux via IC
and bremsstrahlung and to a significant increase of the 
Galactic \diffuse flux below 100~MeV.
For a detailed breakdown of the primary and secondary leptonic components 
as a function of energy see \citet{Porter08} and \citet{Strong10}.
Secondary positrons and electrons are, therefore, indirectly traced by hard
X-rays and \gray{s}. 
The spectrum of
secondary positrons and electrons depends only on the ambient spectrum
of nucleons, the interstellar gas, and the adopted propagation model\footnote{The discovery of enhanced positron fluxes above 10 GeV by the PAMELA instrument is not of importance here since, despite this component, at those energies primary electrons fully dominate the lepton fluxes.}.
Figure~\ref{fig:galdef_id54_z04LMS} show the components of the ISRF that 
contribute to the IC emission in different energy ranges.
The scattering of optical photons provides the majority contribution for 
$\gtrsim 10$~MeV, while the far-infrared dominates in the $\sim 0.1 - 10$~MeV
range, and the CMB is dominant below $\sim 0.1$~MeV.

\subsection{Comparison of SPI spectrum with \GALPROP\ models}

\begin{figure}
\plotone{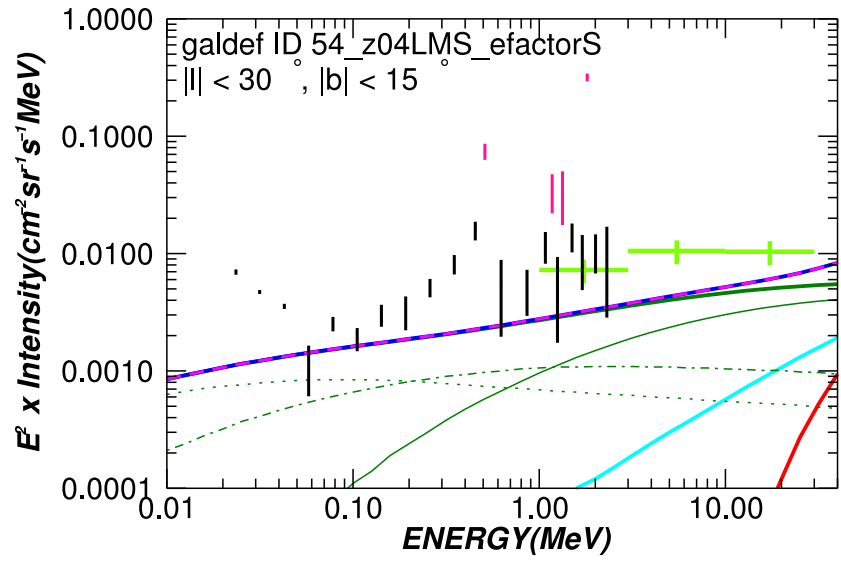}
\caption{Same as Fig.~\ref{fig:galdef_id54_z04LMS}, 
with the primary electron spectrum increased by a factor 2 relatively to 
that used in Fig.~\ref{fig:galdef_id54_z04LMS}. 
 }
\label{fig:galdef_id54_z04LMS_efactorS}
\end{figure}

\begin{figure}
\plotone{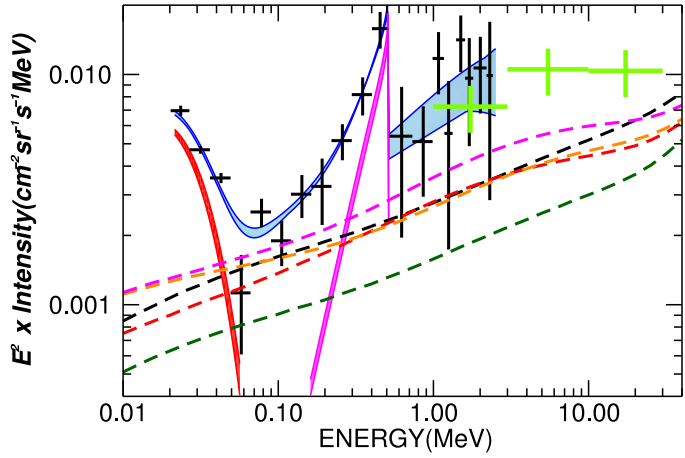}
\caption{
Same as Fig.~\ref{fig:galdef_id54_z04LMS} for different GALPROP configurations.
The dashed  lines are IC models: 
primary electron spectrum of Fig.~\ref{fig:galdef_id54_z04LMS} (dark green), 
primary electron spectrum increased by a factor 2 of Fig.~\ref{fig:galdef_id54_z04LMS_efactorS} (black), increased halo height from 4 kpc to 10 kpc (red) and
increased ISRF in the Galactic bulge (x 10) (orange) and both increased halo height from 4 kpc to 10 kpc and ISRF in the Galactic bulge (x 10) (magenta).
}
\label{fig:galdef_comparison}
\end{figure}

Figure \ref{fig:galdef_id54_z04LMS} compares our baseline GALPROP model 
with the spectrum measured by SPI.
The agreement with the spectral shape is reasonable but the overall intensity
is slightly lower than the data.
Better agreement with the SPI data is obtained by considering a model
with a higher normalisation for the primary electron spectrum, which is 
illustrated in Fig.~\ref{fig:galdef_id54_z04LMS_efactorS} where the 
total electrons are increased by a factor of 2 over the baseline model.
An interpretation for this increase can be that the locally-measured spectrum 
is not typical of the global average at the solar position, or that the CR
source distribution is more peaked toward the inner Galaxy than that used
in the baseline model.
Alternatively, Fig.~\ref{fig:galdef_comparison} shows other 
possibilities to increase the IC component, either by increasing the 
Galactic CR halo height from 4 kpc to 10 kpc, or by increasing the input 
luminosity of the bulge component for the ISRF by a factor of 10.
The uncertainty associated with the bulge input luminosity, metallicity 
gradient, and other factors all contribute so that the ISRF in the inner Galaxy
is not as constrained as observed locally.
Currently, the radial distribution of CR sources is not well known or 
constrained, and a factor 2 increase of the electrons in the inner few kpc 
is possible.
The same effect could be obtained with secondary CR positrons and electrons 
if their CR nuclei progenitors were increased by a similar factor, but then 
this could be inconsistent with the diffuse \gray{} emission measured by
the \FermiLAT\ which is mainly produced by the same CR nuclei.
A larger halo is suggested by analysis of CR data \citep{Trotta10} and 
\gray{s} \citep{Strong2010}.
Unfortunately, the morphology of these components is not
distinguishable by \INTEGRALSPI\ over the spatial region currently covered.
Future observations at higher latitudes would allow discrimination between
the different halo/enhanced ISRF combinations, but this will be the subject
of future work.

\subsection{Comparison of IC spectrum with template-fitted spectrum}
\label{sec:ICspec}

As a further check we compare power-law approximations of 
the \GALPROP\ models with those derived from template fitting.
The best power-law fit to the GALPROP model spectrum shown in 
Fig.~\ref{fig:galdef_id54_z04LMS} (ID 54 z04LMS)
is $N(E)=0.16 \times E^{-1.76}$, where $E$ is the energy in keV. 
The power-law fit to the SPI-extracted IC template is 
$N(E)=0.34 \times E^{-1.79}$, which is two times higher than the \GALPROP\ 
model intensity.
For the model shown in Fig.~\ref{fig:galdef_id54_z04LMS_efactorS} 
(model 54\_z04LMS\_efactorS, which has the primary electrons increased by 
a factor 2), the best power-law fit to the model 
in the range 20~keV--5~MeV is $N(E)=0.29 \times E^{-1.76}$. 
Meanwhile, the power-law fit to the SPI-extracted IC template
is now $N(E)=0.30 \times E^{-1.76}$, which is completely consistent with the
model spectrum.
Similar results can be obtained for the other models involving modifications
to the CR confinement volume, or the intensity of the ISRF in the inner Galaxy.
Thus, it seems that at least some enhancement in the inner Galaxy is required
for the diffuse emission, but it can be due to a combination of effects. 
These we will explore in subsequent work.

\subsection{Electron energies probed by the SPI data}

\begin{figure}
\plotone{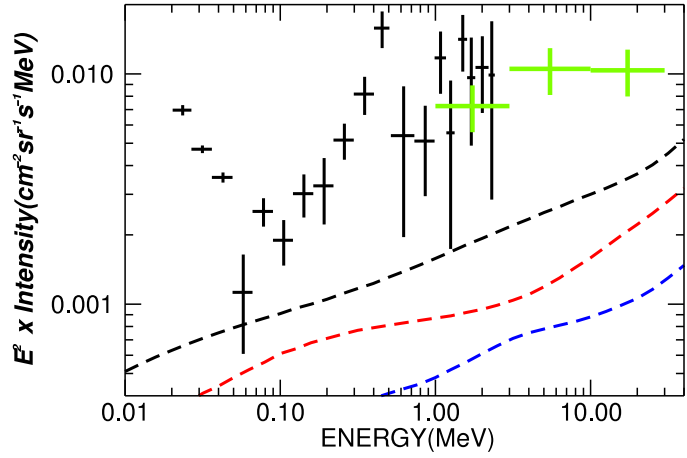}
\caption{
Data and model as in Fig.~\ref{fig:galdef_id54_z04LMS}.
The dashed dotted lines are IC models. Primary electron spectrum 
of Fig.~\ref{fig:galdef_id54_z04LMS} (black), same primary
electron spectrum cutoff below 1~GeV (red), 
and same primary electron spectrum cutoff below 5~GeV (blue). 
}
\label{fig:galdef_electron_cutoff}
\end{figure}

The photon energy range from 50~keV to 2~MeV is sensitive to IC from electrons 
below about 5~GeV. 
To illustrate this, Fig.~\ref{fig:galdef_electron_cutoff}
shows \GALPROP\ model calculations as in 
Fig.~\ref{fig:galdef_id54_z04LMS}, but for primary electron source
spectra cutoff below 1~GeV and 5~GeV, respectively.
The 5~GeV cutoff removes most of the emission in the SPI range, which comes 
mainly from CMB and far-infrared component of the ISRF.
With the 1~GeV cutoff much of the emission is restored, showing that 
most of the IC in the energy range considered in this paper 
comes from electrons between 1 and 5 GeV.  
For these energies, the locally measured CR electron spectrum is strongly
affected by heliospheric modulation and the SPI measurements can allow direct
probing of the {\it interstellar} spectrum of these CRs.
In turn, understanding the heliospheric
transport of CRs could be improved because uncertainties on the true 
interstellar CR spectra directly affect the heliospheric model predictions.

\section{Non-IC \diffuse components :$^{26}$Al, $^{60}$Fe lines and 
annihilation radiation emission}
\label{sec:nonicdiffuse}

The main focus of this paper is continuum emission, so for the 
lines and positronium we restrict ourselves to a check on the 
consistency of our global \diffuse spectrum, because 
these components are extracted simultaneously.
These topics will be explored in greater detail in a separate paper.

The $^{26}$Al line has
been shown to be intrinsically narrow with an 2-$\sigma$ upper limit on 
the width of less than 1.3~keV \citep{Wang09}.
To take into account the emission of this diffuse line, we use an energy band
from 1806 keV to 1812 keV. 
The line is strong compared to the continuum flux in this band and
is essentially unaffected by the assumed underlying power law parameters and
\diffuse continuum spatial shape. 
The contribution in counts of this line through its interaction with the 
detectors (Compton effect, etc.) was subtracted from the continuum prior
to the diffuse continuum data analysis.
The $^{26}$Al line is detected at $\sim 13\sigma$ with a 
flux of $3.3-3.6 \times 10^{-4}$ \phf in the inner Galaxy
($\vert l \vert < 30^{\circ}$, $\vert b \vert < 10^{\circ}$, which agrees 
with earlier measurements \citep[][and references therein]{Wang09}.

The $^{60}$Fe isotope produces two lines at 1173.23 and 1332.50~keV 
detected, depending on the details of their spatial morphology, 
at a level of $\sim 2\sigma$ and $\sim 3\sigma$, respectively. 
Their mean flux in the inner Galaxy is $\sim 6\times10^{-5}$ \phn.
The fluxes derived from this analysis and the $^{60}$Fe to $^{26}$Al ratio of 
$\sim 0.17$ agree with those 
of \citet{Smith04} and \citet{Wang07}. 

The annihilation radiation characteristics have here been measured over a 
wide energy range. 
The derived fluxes are consistent with those in the literature, implying 
a positronium fraction close to 100$\%$, both in the line and the 
positronium continuum. 
The bulge to disk ratio flux is $\sim0.2--0.3$ for the whole Galaxy 
for both the line and the positronium continuum.
For a review on this subject, see \citet{Higdon09}.

\section{Other phenomena in the inner Galaxy}
\subsection{The ``Fermi Bubbles''}
\label{sec:bubbles}

Using \FermiLAT\ data, a claim has been made for two large 
\gray{} ``bubbles'', extending 50$^{\circ}$
above and below the Galactic center, with a width of about 40$^{\circ}$
in longitude \citep{Su10}.
The \gray{} emission associated with these
bubbles appears to have a harder spectrum ($dN/dE \propto E^{-2}$) than the
\gray{s} produced by $\pi^0$-decay by CR nuclei interacting 
with the ISM, or IC emission from CR electrons and positrons modelled using
\GALPROP.
It has also been suggested that the ``bubbles'' are spatially partly 
correlated with the hard-spectrum microwave excess known as 
the \textit{WMAP} haze. 

If the features are real, and are associated, the IC \gray{} emission could 
extend down into the \INTEGRALSPI\ energy range.

We modelled the ``bubbles'' using circular regions centred at $b=30.5^{\circ}$ 
and $b=-30.5^{\circ}$ with a radius of $22^{\circ}$ having a uniform emissivity, 
and tested various combinations of the maps 
(see Table~\ref{table:diffusebestmaps}) along with the ``bubble'' 
templates. 
For the 25--50~keV range, it is not possible to find an unique combination
of maps including the ``bubble'' template that do not result in it 
being assigned a negative coefficient in the fit. 
Imposing the positivity constraint for the fluxes, we find no detection in 
this energy range, as indicated in the table.
Similarly, above 50 keV there is no emission found corresponding to the 
``bubbles''. 
The 2-$\sigma$ upper limits in several
energy bands above 50 keV, assuming an emission with a power-law 
index -2, are given in Table~\ref{table:fermibubbles}.

\subsection{Connection with past activity in the Galaxy}

The increased electron flux in the inner Galaxy could also be related to
past activity in the Galactic Centre, as has been proposed for example for
X-ray flourescence \citep{Terrier10,Capelli11}.
However this would be much more localised near the Galactic Centre, so
that a direct connection seems unlikely. Other more exotic possibilities
for increased electron fluxes, like hypernovae (e.g., related to \gray{}
bursts) cannot be ruled out with the present data.

\section{Summary and discussion}

New results on the diffuse emission spatial morphology and spectrum in 
the range 20~keV--2.4~MeV have been obtained from the analysis of 6 
years of \INTEGRALSPI\ data. 
Over this energy range, what is seen as \diffuse emission is the 
result of the superposition of several physical processes; annihilation 
radiation, cosmic nuclear \gray{} lines, 
diffuse continuum due to interstellar emission, and unresolved sources. 
In the present analysis we have been able to
isolate each of these diffuse components. 
We have explored uncertainties and limitations due to the data reduction 
along with the spatial morphology modelling. 

The \diffuse emission intensity in the central radian 
($\vert l \vert < 30^{\circ}$, $\vert b \vert < 15^{\circ}$) is estimated 
to be one tenth of the total emission (including sources) below 
100~keV and one third in 100--300~keV band. 
The \diffuse emission spectrum has the following main features:

\begin{itemize}

\item The \diffuse continuum spectrum (apart from positronium) is  fitted 
by a power law of index $1.4 - 1.5$, with a flux at 100~keV 
of $1.1 \times 10^{-4}$ \phn. 
This power law, thought to be related to interstellar emission, can be 
decomposed into two spatial components:
the IC component with a power law of index 1.8 and a flux at 100~keV 
of $\sim 10^{-4}$ \phn and another component which can be modelled with the 
an extinction-corrected 4.9 $\mu$m DIRBE-based template, whose spectral shape
is a power law with an index of $\sim 1$ and a flux at 50~keV 
of $\sim 3-4 \times 10^{-5}$ \phn.
This additional component is weak below 200 keV compared to the IC component.

\item  The diffuse continuum flux around 1 MeV is found compatible with 
the \textit{COMPTEL/CGRO} measurement. 

\item The IC emission distribution predicted by the \GALPROP\ code is in 
fair agreement with the data. 
However, a model with an electron spectrum increased by a factor 2 over the 
standard model based on the electrons (plus positrons) 
measured by \FermiLAT\ is in better agreement. 
Also, an increased ISRF in the Galactic bulge or a large Galactic 
CR halo are other reasonable possibilities to that can lead to an 
increased flux. 
The data analysed in this paper do not allow a distinction 
to be made between these possibilities.

\item An additional component is required below 50~keV.
This excess over the IC emission is well modelled with the 
NIR/DIRBE 4.9 $\mu$m map.
This low-energy component has an exponential spectrum with a cutoff 
at 8~keV and a flux at 50~keV of $\sim 2\times 10^{-4}$ \phn.
It can interpreted 
in terms of the stellar origin as proposed by \citet{Krivonos07}.
 
\end{itemize}

The \diffuse continuum emission spectrum obtained with the present analysis
confirms and improves on the results reported in \citet{Bouchet08} 
and \citet{Porter08}.
Below 100~keV, it is also compatible with
results obtained from \INTEGRALIBIS\ analysis \citep{Krivonos07, Turler10}.
In addition, we found that our global \diffuse spectrum is consistent 
through the measured characteristics of nuclear lines ($^{26}$Al, $^{60}$Fe) 
and annihilation radiation spectrum.

There is no detection in the SPI energy range of the ``Fermi bubbles'' 
at the present level of sensitivity.

The \INTEGRALSPI\ will continue to provide new data for at least the next
3 years.
Meanwhile, the \FermiLAT\ will be operating at least over this time span, 
and analysis of its data continue to yield further information on the diffuse 
\gray{} emission $\gtrsim 100$~MeV.
An analysis showing the complementarity of the data provided by 
these instruments enabling us to probe the physical processes producing the 
diffuse emission is given in \cite{Strong2010}.
Together with existing data from other instruments, the data from these two 
currently operating missions allows the investigation of the CR
electron spectrum at all energies, which will eventually enable 
an unambiguous decomposition of the diffuse \gray{} sky.

\acknowledgments
The \INTEGRALSPI\ project has been completed under the responsibility and
leadership of CNES.
We are grateful to ASI, CEA, CNES, DLR, ESA, INTA, NASA and OSTC for support.

\GALPROP\ development is partially funded via NASA grants NNX09AC15G and 
NNX10AE78G.

Some of the results in this paper have been derived using the HEALPix 
\citep{healpix} package.

\newpage
\appendix
\section{Form of INTEGRAL/SPI the response matrix}
\label{appendixA}

The SPI response R has been splitted into 3 components which
correspond to the detector response to the photopeak events: R$^{(1)}$,
non-photopeak events that interact first in a detector:
R$^{(2)}$(Compton interaction,\ldots{}) and photons that interact first
in the passive material surrounded the detector: R$^{(3)}$.

\begin{equation}
R(E,E_{ph},\theta,d)=\sum_{i=1}^{3}R^{(i)}(E,E_{ph},\theta,d)
\end{equation}

Here E stands for the detected energy, E$^{ph}$ for the incident
photon energy, $\theta$ for the incident photon direction relative
to the telescope axis and d the detector number.
The templates $R^{(i)}$ were found to a good approximation to not vary with
detector or incident photon direction, only their normalization changes
\citep{Sturner03}. The templates are thus determined for a given
photon energy and only the
normalization or efficiency is calculated for every incident photon
direction and detector number.
In short,  each component has been again splitted in two
parts. The first part is Imaging Response Function (IRF) and the
second part is the Redistribution Matrix File (RMF). The IRFs contain the
detector effective area as the function of the detector number (d), the
incident photon direction ($\theta$) and the incident photon energy
(E$_{ph}$). The RMFs contain information about the energy distribution
of detector counts (E) for photons of a given incident energy
(E$_{ph}$). The response is rewritten as:

\begin{eqnarray}
R(E,E_{ph},\theta,d)&=&\sum_{i=1}^{3} IRF^{(i)}(E_{ph},\theta,d) \times RMF^{(i)}(E,E_{ph}) \\
& &  \int R(E,E_{ph}) dE= 1 
\end{eqnarray}

This decomposition into static components reduces the computation time
and storage. More detailed can found in Sturner at al., 2003.

The detector counts C produces  by a source (incident photon direction $\theta$) emitting a spectrum S(E$_{ph}$,$\theta$) is

\begin{eqnarray}
C(E,d)&=&\sum_{i=1}^{3} \int IRF^{(i)}(E_{ph},\theta,d) \times RMF^{(i)}(E,E_{ph}) S(E_{ph},\theta) dE_{ph} \\
&=& \sum_{i=1}^{3} C^{(i)}(E,\theta,d) 
\end{eqnarray}

For N$_p$ exposures (or pointings), N$_d$ detectors, N$_s$ sources and N$_e$ energy band (Sec. 3), the equation is

\begin{equation}
C(E,d,p) = \sum_{s=1}^{N_s}  \sum_{i=1}^{3} \int IRF^{(i)}(E_{ph},\theta_s,d,s) \times RMF^{(i)}(E,E_{ph}) S(E_{ph},\theta_s)) dE_{ph} \\
\end{equation}
$\theta_s$ is the direction of the source number s relative to the telescope axis and S(E$_{ph}$,$\theta_s$) its incident photon spectrum.

The number of equations to hold simultaneously is then $ N_d \times N_p \times N_e$ and the number of unknowns is $N_s \times N_e$,
assuming no background and constant flux sources (Sec. 3).

Taking into account completely the response matrix requires to solve a
more complex equation than eqn 1 with  extra additional dimensions in
both data space (all detected energy or data channels)
and photon space (incident photon energy).
Fortunately, it is possible to use a close approximation of these
equations to reduce it to the form of eq. 1.
The latter system of equation is subsequently solved several times (for each energy band), with 
$N_d \times N_p$ equations to hold simultaneously and $N_s$ unknowns.

\subsection{Flux extraction in counts space}

To reduce the dimension of the problem to be solved, a
flux extraction in counts space is first done and the resulting source
counts are then converted into incident photon spectrum. In this case
only the photopeak response part is used.

\begin{equation}
C(E,d)= C^{(1)}(E,\theta,d) \times (1+\frac{C^{(2)}(E,\theta,d)+C^{(3)}(E,\theta,d)}   {C^{(1)}(E,\theta,d)}
\end{equation}

The term IRF$^{(1)}$ is is photopeak efficiency (and omitting the energy resolution term for simplicity), 
\begin{eqnarray}
RMF^{(1)}(E,E_{ph}) &=& \delta(E,E_{ph}) \\
C^{(1)}(E,\theta,d) &=& IRF^{(1)}(E,d) \times S(E,\theta)
\end{eqnarray}

Then 
\begin{equation}
C(E,d)=IRF^{(1)}(E,d) \times S(E,\theta) \times \beta(E,\theta,d)
\end{equation}

For a fixed energy E, making the approximation that  whatever the event types, 
the spatial distribution of counts over the detector plane (in function of the detector number d) is similar and
hence does not depend on the detector number d ($\beta(E,\theta,d) \simeq \beta(E,\theta)$) , then

\begin{equation}
C(E,d)=IRF^{(1)}(E,\theta,d) \times S(E,\theta) 
\times \beta(E,\theta)= IRF^{(1)}(E,\theta,d) \times S^{counts}(E,\theta)
\end{equation}

Here $S^{counts}$ is called the flux in counts space.

\subsection{Direct photon flux extraction \textendash{}
\textquotedblleft{}pseudo-efficiency\textquotedblright{}}

It is possible to extract the sources photon spectrum directly
from the data. For that the emission spectrum spectral shape is
assumed to be known and to be continuous in energy. This shape can for
example, be extracted after a first flux extraction in counts space and
conversion into incident photon spectrum. $S^{Fitted}$ being the fitted photon spectrum, then

\begin{equation}
C(E,d)= \frac {\sum_{i=1}^{3} \int IRF^{(i)}(E_{ph},\theta,d) \times RMF^{(i)}(E,E_{ph}) S^{Fitted}(E_{ph},\theta) dE_{ph} } 
        {S^{Fitted}(E,\theta)}  \times S(E,\theta)
\end{equation}

If $S^{Fitted}(E,\theta)$ is suffuciently close to the incident spectrum $S(E,\theta)$ then the above formula
predicts counts in data space. It can can be rewritten

\begin{equation}
C(E,d)=R^{Pseudo}(E,\theta,d) \times S(E,\theta)
\end{equation}
\\
$R^{Pseudo}$ is the response of the instrument assuming a continuous photon emission spectrum of known spectral shape
(for 1 incident photon in each energy band).

\subsection{'Simplified' system of equations}

Finally, for  $N_s$ sources located in the field of view,
the data D(d,p,E)obtained during an exposure (pointing p) in the detector d for given energy
E is

\begin{equation}
D_{dp}=\sum_{j=1}^{N_s} R_{dp,j} S_{p,j} + B_{dp}
\end{equation}

$B_{dp}$ is the background obtained during an exposure (pointing p) in
detector d for given energy E. R can be be either $IRF^{(1)} or R^{Pseudo}$. The equation will be solve in this latter form
(eq. 1).

If the sources emission spectrum follows a known spectral shape, then
the counts predicted by the pseudo efficiency method is more realistic.
Anyway this method can be more sensitive to errors in the simulated
response matrix and storage simplification made, 
and requires the knowledge of the incident photon
spectrum which might be continuous in energy.
In the other hand, the flux extraction in counts space, may predict inaccurate source
counts as the photopeak like assumed response is not the true response.

\section{Impact of the energy redistribution matrix form on \diffuse measurement}

The first approximation (counts flux approximation) used only the photopeak part of the response
to obtain fluxes in counts space in a first step. Then,
these counts fluxes are then converted into incident photon spectrum.
The \textquotedblleft{}pseudo-efficiency\textquotedblright{} response allows
obtaining directly an incident photon spectrum.

\begin{figure}
\plotone{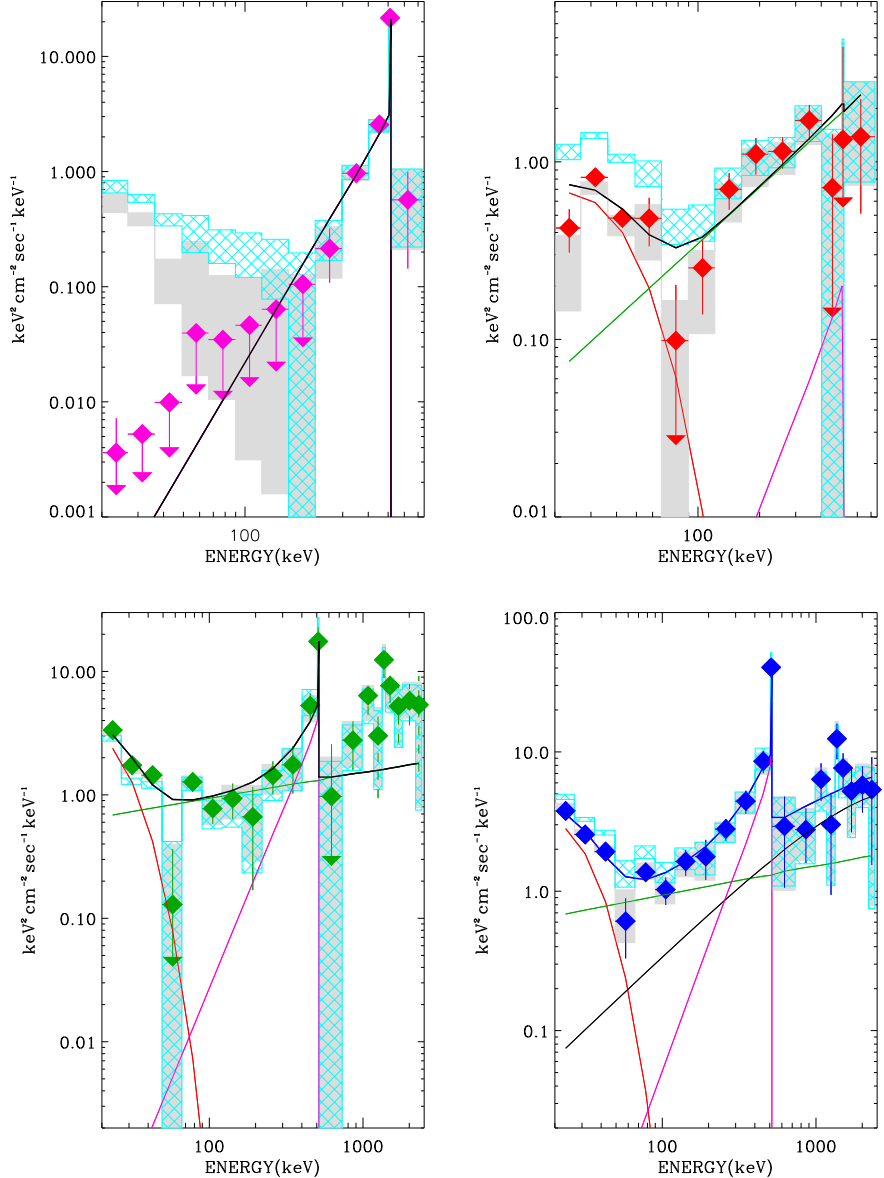}
\caption{
Spectrum of the diffuse emission for
$\vert l \vert < 30^{\circ}$ and $\vert b \vert < 15^{\circ}$ for 
each extracted components. 
For each plot, the 1-$\sigma$ envelope obtained with the counts flux matrix, 
both for
sources and diffuse, is shown as the cyan hatched region,  
while the spectrum that is obtained with the  
pseudo-efficiency for sources and counts flux matrix for the diffuse emission 
is shown as a grey-shaded region.
The crosses and diamond symbols correspond to the components
obtained with the pseudo-efficiency matrix for both sources and diffuse.  
The black line is the resulting emission spectrum.
Upper left: magenta: annihilation radiation spectrum (line + positronium)
Upper right: red: 
emission of \textquotedblleft{}unresolved\textquotedblright{} sources,
mainly cataclysmic variable and coronally active stars.
Lower left: green line is the  deduced interstellar  emission.
Lower right: the green line is the \GALPROP\ IC power-law fitted on c), 
the red is the A$_{4.9 \mu {\rm m}}$ power-law fitted on b), 
magenta is the deduced total annihilation
radiation, blue points with the diamond symbol are the total diffuse emission 
while the blue continuous line is its best fitted model (sum of all 
diffuse fitted components).}
\label{fig:diffuse_components_details}
\end{figure}

For the counts flux approximation, for each point-source a small
fraction of the non-photopeak events could have been included in the
background or vice-versa. For a given point source, the effect is negligible on the extracted counts spectrum.
Nevertheless, for the present analysis
and due to the numerous point sources (accumulation effect) and their spatial distribution
(most of them located around the Galaxy plane), these counts residuals 
may sum-up to mimic a false \diffuse emission like supplementary component.
This effect is visible in fig.~\ref{fig:diffuse_components_details}, actually it is negligible for the 
whole sky spectrum but is important for the central radian region.
It is not the case for the \textquotedblleft{}pseudo-efficiency\textquotedblright{} approximation
as long as a close estimate of each point-source incident photon spectral
emission shape is available. 
The method gives by construction a better
extraction of sources counts and separation from other component.
Nevertheless, the method is sensitive to errors in the non-photopeak
events response matrices. \\
Figure \ref{fig:diffuse_components_details} 
shows the resulting diffuse
spectrum of each components and the the comibined spectrum for the two
approximations mentioned above plus a third case where the
\textquotedblleft{}pseudo-efficiency\textquotedblright{} is used for
point-sources, while photopeak response is used for all the diffuse
components. These three analyses were done in parallel at each step of the
study. They give an idea of the uncertainties/systematic.
All produces compatible results above 100 keV, below a larger
intensity is found for the diffuse components as suspected. Anyway, we
use results obtain with
\textquotedblleft{}pseudo-efficiency\textquotedblright{} as it is
better adapted to model counts of all sources in the data space and
hence a better separation of the background and sources terms.

\newpage

\newpage

\newpage

\newpage

\newpage
\clearpage

\newpage
\clearpage

\newpage
\clearpage

\newpage
\clearpage

\newpage
\clearpage

\newpage
\clearpage

\newpage
\clearpage

\newpage
\clearpage
\begin{deluxetable}{lcccccc}
\tablewidth{0pt}
\tablecaption{Information on exposures }
\tabletypesize{\scriptsize}
\tablehead{
\colhead{Energy band (keV)}
&\colhead{  27--36 }
&\colhead{  25--50 }
&\colhead{ 50--100 }
&\colhead{ 100--200 }
&\colhead{ 200--600 }
& \colhead{600--1800 }
}
\startdata
Number of exposures                       &  38699 &  38699 &  38699 &  38699 &  38699 &  36486     \\
Reduced $\chi^2$                          & 1.18   & 1.27   & 1.07   & 1.05   & 1.04   & 1.03      \\
Degree of freedom                         & 656914 & 649991 & 664354 & 666411 & 666595 & 628753    \\
Number of point-sources                   & 256    & 254    &  121   & 53     &  26    & 4         \\
Number of parameters                      & 15581  &  22504 &  8141  &   6084 &  5900  &  5606    \\
Exposure maximum reduced $\chi^2$(p)           &   4.2  &   5.2  &    3.1 &    3.0 &   3.0  & 3.1       \\
\multicolumn{1}{l}{Exposures with}        &        &        &        &        &        &        \\
\multicolumn{1}{c}{Reduced $\chi^2$(p)$~< 2$}  & 38342  & 38690  & 38436  & 38471  & 38473  & 36450  \\
\multicolumn{1}{c}{Reduced $\chi^2$(p)$~< 3$}  & 38513  & 38232  & 38698  & 38698  & 38699  & 36485  \\
\enddata
\tablecomments{Reduced $\chi^2$  is the total chi-square between the data and the best sky model convolved with the instrument response  divided by the  degree of freedom. Reduced $\chi^2$(p) is the  exposure chi-square divided by the number of working detectors (computed for each exposure p).}
\label{table:exposuresinfolarges}
\end{deluxetable}

\begin{deluxetable}{lcccccccc}
\tablewidth{0pt}
\tablecaption{Information on spectral energy bands }
\tabletypesize{\scriptsize}
\tablehead{
\colhead{Energy band (keV)}
&\colhead{ 20--27 }   
&\colhead{ 27--36 }
&\colhead{ 36--49 }
&\colhead{ 49--66 }
&\colhead{  66--90 }
&\colhead{  90--121 }
&\colhead{ 121--163 }
&\colhead{ 163--220 }
}
\startdata
Number of point sources     & 256      &    256   &     251  &    193   &     158  &    122   &   77     &  48      \\
Reduced $\chi^2$            & 1.088400 & 1.178813 & 1.124047 & 1.058390 & 1.087771 & 1.031193 & 1.024928 & 1.025746 \\
Degree of freedom           &  656914  &  656914  &  656921  &  664753  &  664804  &  666291  &  666370  &  666415  \\
\\
\hline 
Energy band (keV)           & 220-298  & 298-402  & 402-505  & 505-516  & 516-732  & 732-988  & 988-1170  & 1170-1176 \\
\hline
Number of point-sources     &   34     &   17     &   13     &   11     & 11       & 11       &   2       &     2     \\
Reduced $\chi^2$            & 1.007953 & 1.006851 & 1.013379 & 0.999983 & 1.014998 & 1.002269 & 1.003606  & 1.000524  \\
Degree of freedom           &  666587  &  666604  &  666608  &  666610  &  666610  & 628746   &  628755   &  628755   \\
\\
\hline
Energy band (keV)           & 1176-1330 & 1330-1336 & 1336-1400 & 1400-1600 & 1600-1806  & 1806-1812 & 1812-2200 & 2200-2414 \\
\hline
Number of point-sources     &     2     &     2     &  2        &    2      &   2        &   1       &     1     &   1       \\
Reduced $\chi^2$            & 1.000548  & 0.997189  & 0.999347  & 1.004202  & 1.003443   & 1.000153  & 1.002443  & 1.001612  \\
Degree of freedom           &  628755   &  628755   &  628755   &  628755   &  628755    &  628756   &  628756   &  628756   \\
\\
\enddata
\tablecomments{PSD+PE data (38699 exposures) are used for energies below 650 keV and only PSD data (36486 exposures) above.}
\label{table:exposuresinfospectral}
\end{deluxetable}

\begin{deluxetable}{lccccccc}
\tablewidth{0pt}
\tablecaption{Galactic \diffuse morphology modelling with synthetic maps }
\tabletypesize{\scriptsize}
\tablehead{
\multicolumn{8}{c}{\textbf{Modelling with a single synthetic map}}  \\
\hline \hline
\colhead{Energy band (keV)} & \colhead{25--50}   & \colhead{50--100}    & \colhead{100--200} &
\colhead{200--600}    & \colhead{600--1200}   & \colhead{1200--1800}    & \colhead{600--1800}}
\startdata
\bf{\emph{1.25$\mu$m}} &       345.1 &        84.5 &       119.7 &        74.4 & \bf{\emph{0.1}} &\bf{\emph{0.6}} &\bf{\emph{0.5}}  \\
2.2$\mu$m              &       181.9 &        65.7 &        85.6 &        47.0 &         3.0 &         1.3 &         4.5  \\
3.5$\mu$m              &       153.7 &        55.0 &        76.6 &        45.3 &         5.4 &         1.6 &         7.4  \\
4.9$\mu$m              &       160.3 &        46.8 &        75.0 &        55.8 &         7.7 &         1.8 &        10.1  \\
12$\mu$m               &       961.6 &        90.4 &       109.6 &        78.8 &         8.6 &         4.3 &        12.6  \\
25$\mu$m               &      1149.7 &        85.3 &       127.4 &        86.5 &        12.9 &         6.5 &        19.3  \\
60$\mu$m               &      1000.0 &        54.7 &        82.8 &        73.6 &        14.4 &         5.8 &        20.4  \\
100$\mu$m              &       850.9 &        50.7 &        76.0 &        68.1 &        11.5 &         4.9 &        16.2  \\
140$\mu$m              &       863.1 &        56.1 &        86.9 &        74.2 &        10.6 &         5.0 &        15.2  \\
240$\mu$m              &       952.9 &        74.9 &       111.6 &        88.1 &        10.4 &         5.2 &        15.3  \\
$^a$A$_{2.2\mu{\rm m}}$      &       574.2 &        75.4 &       101.3 &        50.7 &         5.5 &         2.3 &         7.1  \\
$^a$A$_{3.5\mu {\rm m}}$      &       483.6 &        51.5 &        69.8 &        37.8 &         6.0 &         2.9 &         8.0  \\
$^a$A$_{4.9\mu {\rm m}}$      &       447.8 &        29.1 &        71.0 &        45.8 &         7.6 &         4.9 &        11.3  \\
$^a$A$_{12\mu {\rm m}}$       &       647.0 &        31.0 &        82.7 &        54.5 &         9.6 &         6.3 &        14.6  \\
CO                    &       838.5 &        76.2 &       108.5 &        96.1 &         9.6 &         9.6 &        17.8  \\
HI                    &      1522.3 &       182.4 &       347.8 &       222.6 &        20.3 &        10.7 &        32.1  \\
IC-ID54z04LMS         &       219.9 &        29.3 &        44.9 &        18.8 &         2.1 &         1.1 &         2.4  \\
\bf{IC-ID54z04LMS-efactor} &       231.0 &        31.9 &        48.1 &    18.9 & \bf{2.1}    &    \bf{1.1} &    \bf{2.4} \\
Degree of freedom     &    649992   &    664355   &    666412   &    666596   &    628747   &    628753   &    628753    \\
\hline
\\
\multicolumn {8}{c}{\textbf{Modelling with a combination of IC-ID54z04LMS-efactor (hereafter IC) plus another synthetic map}} \\
\hline \hline
Energy band (keV)      &       25-50 &      50-100 &     100-200 &     200-600 & 600-1200    & 1200-1800   & 600-1800     \\
\hline
\bf{\emph{IC + 1.25 $\mu$m}}&       180.8 &        31.9 &        47.8 &        18.9 &\bf{\emph{0.0}}&\bf{\emph{0.4}}&\bf{\emph{0.1}}  \\
IC + 2.2 $\mu$m          &        59.0 &        30.2 &        33.1 &        17.3 &         1.5 &         0.7 &         1.6  \\
IC + 3.5 $\mu$m          &        23.8 &        25.6 &        25.7 &        14.3 &         2.0 &         0.7 &         2.1  \\
\bf{\emph{IC + 4.9 $\mu$m}}& \bf{\emph{9.1}} &        20.0 &        20.8 &        12.5 &         2.1 &         0.6 &         2.3  \\
IC + 12 $\mu$m           &       215.3 &        27.7 &        25.7 &        12.7 &         2.1 &         1.1 &         2.4  \\
IC + 25 $\mu$m           &       203.4 &        16.1 &        13.3 &         4.8 &         2.1 &         1.1 &         2.4  \\
\bf{\emph{IC + 60 $\mu$m}}&       196.2 &\bf{\emph{5.2}}&\bf{\emph{3.8}}&\bf{\emph{0.1}} &         2.1 &         1.1 &         2.4  \\
IC + 100 $\mu$m          &       188.8 &         9.9 &        10.9 &         5.6 &         2.1 &         1.1 &         2.4  \\
IC + 140$\mu$m          &       193.7 &        13.2 &        17.0 &         9.0 &         2.1 &         1.1 &         2.4  \\
IC + 240$\mu$m          &       215.4 &        21.7 &        27.2 &        13.0 &         2.1 &         1.1 &         2.4  \\
IC + $^a$A$_{2.2\mu m}$  &       231.0 &        31.9 &        48.1 &        18.8 &         2.1 &         1.1 &         2.4  \\
IC + $^a$A$_{3.5\mu m}$  &       221.7 &        29.7 &        39.9 &        15.3 &         2.1 &         1.1 &         2.4  \\
\bf{IC + $^a$A$_{4.9\mu m}$} &  \bf{101.3} &    \bf{6.6} &   \bf{20.5} &    \bf{7.0} &         2.1 &         1.1 &         2.4  \\
IC + $^a$A$_{12\mu m}$   &       145.6 &         3.8 &        20.8 &         7.5 &         2.1 &         1.1 &         2.4  \\
IC + CO                &       120.8 &        15.0 &        13.3 &        11.8 &         2.1 &         1.1 &         2.4  \\
IC + HI                &       231.0 &        31.9 &        48.1 &        18.9 &         2.1 &         0.8 &         2.4  \\
Degree of freedom      & 649991      &    664354   &    666411   &    666595   &    628746   &    628752   &    628752    \\
\hline
\\
\multicolumn {8}{c}{\textbf{Modelling with a combination IC plus all other (non IC) synthetic map}} \\
\hline \hline
Energy band (keV)      &       25--50 &      50--100 &     100--200 &     200--600 & 600--1200    & 1200--1800   & 600--1800     \\
\hline
All maps$^*$           &         0.0 &         0.0 &         0.0 &         0.0 &         0.0 &         0.0 &         0.0  \\
Degree of freedom      & 649976      &    664339   &    666396   &    666580   &    628731   &    628737   &    628737    \\
\enddata
\tablecomments{Results of template fitting in several energy ranges. 
Values of the maximum-likelihood ratio $-2 \times (Ln~L - ln~L_{0}) \approx \Delta \chi^2$.
The sky model consists of point sources + diffuse (maps) + bulge 
(Gaussian of $3.2^{\circ}$ and $11.8^{\circ}$ centered at l=$-0.6^{\circ}$ and b=$0.0^{\circ}$ for 
energies below 516 keV, except for the "all map" case (*) where the bulge is included for all energy bands.
First column contains the synthetic map name used, first line dispalys the energy bands in  keV unit.
Note that the maximimum-likelihood ratio is zero by definition for the ``all  maps'' 
model ($^*$), which corresponds to the combination of the best models by band.
($^a$) are IR  extinction-corrected maps (Sec.~\ref{sec:ridgespatial}).\\ 
IC-ID54z04LMS and IC-ID54z04LMS-efactor
are IC distribution maps computed with GALPROP code
(Sec.~\ref{sec:galprop}). 
For each energy band, highlighted numbers and models are the absolute best fit model(bold-italic)
and the  model used throughout this analysis.
The reduced $\chi^2$, for the model in bold, are 1.27, 1.07, 1.05, 1.04, 1.03, 1.01 and 1.02 respectively for the 25--50, 50--100, 100--200, 200--600, 600--1200, 1200--1800, and 600--1800~keV bands.
}
\label{table:diffusebestmaps} 
\end{deluxetable}

\begin{deluxetable}{lll}
\tablewidth{0pt}
\tablecaption{Diffuse ($\vert l \vert < 30^{\circ}$,$\vert b \vert < 15^{\circ}$) spectral fit }
\tabletypesize{\scriptsize}
\tablehead{
\colhead{Spectral model} & \colhead{ Parameter}   & \colhead{Value} }
\startdata
IC power law               &    index        & 1.79 (fixed)                              \\
                           & Flux \factf~at 100 keV & $0.92$  \phn (fixed) \\
\\
A$_{4.9 \mu{\rm m}}$ power law    & index           & 0.95 (fixed)                                 \\
plus cutoff                & Cutoff energy   & $3411_{-1170}^{+2371}$  keV                            \\
                           & Flux \factf~at 100 keV & $0.34 $ \phn (fixed) \\ 
\\
exponential cutoff         & Cutoff energy   & $7.7 \pm 0.1          [_{-0.6}^{+0.7}]$  keV                            \\
                           & Flux \factf~at 50 keV  & $1.60 \pm 0.06 [_{-0.35}^{+0.40}]$ \phn  \\ 
\\
Positronium continuum      & Flux \factf           & $67.3\pm 14.6   [\pm 22.4]$  \phf  \\ 
Gaussian line at 511 keV   & Flux \factf           & $15.8 \pm 2.7 [\pm 4.1]$  \phf \\
$\chi^2$ (dof)             & 23(16)          &                                                          \\
\\
\multicolumn {3}{c}{\textbf{Fit of each of the extracted spatial component separately}} \\
\hline \hline
\multicolumn {3}{c}{\textbf{Bulge}$^a$} \\
\hline
Positronium contiuum      &  Flux \factf      & $28.9 \pm 2.9   $ \phf     \\ 
Gaussian line at 511~keV  &  Flux \factf      & $9.1 \pm 0.7    $ \phf   \\
$\chi^2$ (dof)            &  11(11)     &                                                      \\ 
\\
\multicolumn {3}{c}{\textbf{Extracted A$_{4.9 \mu{\rm m}}$ like spatial morphology component}$^b$} \\
\hline
Power law                &  index          & $0.95_{-0.04}^{+0.02}         [_{-0.33}^{+0.27}]$   \\ 
                         & Flux \factf~at 100 keV & $0.34_{-0.04}^{+0.08}  [_{-0.09}^{+0.21}]$ \phn \\
Exponential cutoff       & Cutoff energy   & $11 \pm 1                     [_{-3}^{+6}]$ keV                   \\
                         & Flux at 50 keV  & $1.07_{-0.25}^{+0.22}         [_{-0.59}^{+0.62}]$ \phn \\ 
Positronium continuum    &  Flux \factf          & $1.8_{-1.8}^{+15.8}$ \phn     \\ 
Gaussian line at 511 keV & Flux \factf           & $0.0_{-0.0}^{+1.2}$ \phn   \\
$\chi^2$ (dof)           & 11(13)          &                                \\
\\
\multicolumn {3}{c}{\textbf{Extracted IC spatial morphology component}$^{c}$} \\
\hline
Power law                &  index          & $1.79_{-0.04}^{+0.03}       [_{-0.25}^{+0.25}]$            \\
                         & Flux \factf~at 100 keV & $0.92_{-0.12}^{+0.17} [_{-0.17}^{+0.12}]$ \phn \\ 
Eponential cutoff        & Cutoff energy    & $6.3_{-0.3}^{+0.5}         [_{-2.1}^{+3.1}]$  keV                            \\
                         & Flux \factf~at 50 keV & $0.61_{-0.12}^{+0.28} [_{-0.51}^{+1.12}]$ \phn \\ 
Positronium continuum    & Flux \factf  & $35.7_{-13.7}^{+21.0}   $ \phf         \\ 
Gaussian line at 511 keV & Flux \factf  & $6.8_{-2.6}^{+4.1}$ \phf  \\
$\chi^2$ (dof)           & 10(8) &                                      \\
\enddata
\tablecomments{Spectral model fitting results using both spatial nmorphology and spectral decomposition information.
For the overal fit (20 keV--2.4 MeV), channels containing $^{26}$Al and  $^{60}$Fe lines are omitted,
the annihilation line energy is fixed to 511~keV with a FWHM fixed to 2.5 keV.
The separate component fitting results (a-c) applied to fix some parameters and to enable a multi-component fit rely on the data below 1 MeV.
The quoted errors are for a single parameter of interest ($\chi^2=\chi^2_{minimum}$+1.0) except for
those indicated between parenthesis that are for a single spectral model and  2 free parameters simultaneously
($\chi^2=\chi^2_{minimum}$+2.35).}
\label{table:diffusebestfits}
\end{deluxetable} 

\begin{deluxetable}{lccccc}
\tablewidth{0pt}
\tablecaption{``Fermi bubbles'' 2-$\sigma$ upper limits (Fluxes are in unit of $10^{-3}$ \phf) }
\tablehead{
\colhead{Energy band (keV)} 
&\colhead{25--50$^*$}   
&\colhead{50--100}  
&\colhead{100--200} 
&\colhead{200--600}  
&\colhead{600--1800} }
\startdata
Upper Galactic latitude bubble    & $< 1.6 $ & $< 1.8 $ & $< 1.3 $ & $< 1.5 $ & $< 1.2 $ \\
Lower Galactic latitude bubble    & $< 1.6 $ & $< 1.8 $ & $< 1.3 $ & $< 1.5 $ & $< 1.2 $ \\
\enddata
\tablecomments{$^*$For the 25-50 keV band (Sec.~\ref{sec:bubbles}), the fluxes of all the sky components are constrained to be positive.}
\label{table:fermibubbles}
\end{deluxetable}


\newpage

\begin{thebibliography}{46}

\expandafter\ifx\csname natexlab\endcsname\relax\def\natexlab#1{#1}\fi

\bibitem[Abdo et al.(2009)]{Abdo09}        
Abdo, A.~A., Ackermann, M., Ajello, M., et al., \ 2009, \prl, 103, 25, id. 251101

\bibitem[Abdo et al.(2010)]{Abdo10}        
Abdo, A.~A., Ackermann, M., Ajello, M., et al., \ 2010, \prl, 104, 10, id. 10110

\bibitem[Amestoy et al.(2006)]{Amestoy06}
Amestoy, P. R., Guermouche, A., L'Excellent, J. Y., \& Pralet, S. \ 2006, Parallel 
Computing, Vol 32 (2), pp 136-156.

\bibitem[Bleach et al.(1972)]{Bleach72}        
Bleach, R.~D., Boldt, E.~A., Holt, S.~S., Schwartz, D.~A, et al., \ 1972, \apj, 174, L101

\bibitem[Bird et al.(2010)]{Bird10}
Bird, A.~J., Bazzano, A., Bassani, L., et al., \ 2010, \apjs, 186, 1

\bibitem[Bouchet et al.(2005)]{Bouchet05}        
Bouchet, L., Roques, J. P., Mandrou, P., et al., \ 2005, \apj, 635, 1103

\bibitem[Bouchet et al.(2008)]{Bouchet08}        
Bouchet, L., Jourdain, E., Roques, J.~P, et al., \ 2008, \apj, 679, 1315

\bibitem[Bouchet et al(2009)]{Bouchet_otranto09}        
Bouchet, L., Jourdain, E., Roques, J.~P.,  et al., \ 2009, 
in Proceedings of The Extreme sky: Sampling the Universe above 10 keV, http://pos.sissa.it/cgi-bin/reader/conf.cgi?confid=96, p.16

\bibitem[Bouchet et al.(2010)]{Bouchet10}        
Bouchet, L., Roques, J.~P., \& Jourdain, E., \ 2010, \apj, 720, 1772

\bibitem[Capelli et al.(2011)]{Capelli11}   
Capelli R., Warwick R.~S., Porquet D., et al., \ 2011, \aap, 430, A38

\bibitem[Churazov et al.(2010)]{Churazov10}
Churazov, E., Sazonov, S., Ysygankov S., et al., \ 2010, \mnras,  in press (arXiv:1010.0864C)

\bibitem[Dame, Hartmann \& Thaddeus(2001)]{Dame01}
Dame, T.~M., Hartmann, D., \& Thaddeus, P., \ 2001, \apj, 547, 792

\bibitem[Dickey \& Lockman(1990)]{Dickey90}
Dickey, J.~M., \&  Lockman, F.~J., \ 1990, ARA\&A , 28, 215

\bibitem[Diehl et al.(1995)]{Diehl95}        
Diehl, R., Dupraz, C., Bennett, K., et al.,  \ 1995, \aap, 298, 445

\bibitem[Diehl et al.(2006)]{Diehl06}        
Diehl, R., Halloin, H., Kretschmer, K., et al., \ 2006, \nat, 439, 435

\bibitem[Dogiel et al.(2002)]{Dogiel02}
Dogiel, V.~A.,  Sch\"onfelder, V., \& Strong, A.~W., \ 2002 \aap, 382,730 

\bibitem[Dubath et al.(2005)]{Dubath05}
Dubath, P., Kn\"odlseder, J., Skinner, G.~K., et al., \ 2005, \mnras, 357, 420

\bibitem[Ebisawa et al.(2005)]{Ebisawa05}        
Ebisawa, K., Tsujimoto, M., Paizis, A., et al., \ 2005, \apj, 635, 214

\bibitem[G\'{o}rski et~al.(2005)]{healpix}
G\'{o}rski, K.~M., et~al., \ 2005, \apj, 622, 759

\bibitem[Hands et al.(2004)]{Hands04}        
Hands, A.~D.~P., Warwick, R.~S., Watson, M.~G., \& Helfand, D.~J. \ 2004, \mnras, 351, 31

\bibitem[Harris et al.(2005)]{Harris05}  
Harris, M.~J, Kn\"odlseder, J., Jean, P., et al., \ 2005, \aap, 433, L49

\bibitem[Higdon, Lingenfelter \& Rothschild(2009)]{Higdon09}
Higdon, J.~C., Lingenfelter, R.~E., \& Rothschild,R.~E., \ 2009, \apj, 698, 350

\bibitem[Jensen et al.(2003)]{Jensen03}
Jensen, P.~L., Clausen, K., Cassi, C., et al., \ 2003, \aap, 411, L7

\bibitem[Jourdain \& Roques(2009)]{JR2009}
Jourdain, E. \& Roques J.~P., \ 2009, \apj, 704, 17


\bibitem[Kaneda et al.(1997)]{Kaneda97}        
Kaneda, H., Makishima, K., Yamauchi, S., et al., \ 1997, \apj, 491, 638

\bibitem[Krivonos et al.(2007)]{Krivonos07}        
Krivonos, R., Revnivtsev, M., Churazov, E., et al., \ 2007, \aap, 463, 957
        
\bibitem[Lebrun et al.(2004)]{Lebrun04}        
Lebrun, F., Terrier, R., Bazzano, A., et al., \ 2004, \nat, 428, 293

\bibitem[Mahoney, Jacobson \& Lingenfelter(1982)]{Mahoney82}        
Mahoney, W.~A., Ling J.~C, Jacobson, A.~S., \& Lingenfelter, R.~E. \ 1982, \apj ,.262, 742

\bibitem[Ore \& Powell(1949)]{Ore49}
Ore, A. \& Powell, J. \ 1949, Phys. Rev., 75, 1696

\bibitem[Pl\"uschke et al(2001)]{Pluschke01}        
Pl\"uschke, S., Diehl, R., Sch\"onfelder, V., et al., \ 2001,  ESA-SP 459, 55

\bibitem[Porter et al.(2008)]{Porter08}                
Porter, T.~A., Moskalenko, I.~V., Strong, A.~W., et al., \ 2008, \apj, 682, 400

\bibitem[Rouet(2009)]{Rouet09}
Rouet, F.~R. \ 2009, Partial computation of the inverse of a large sparse matrix-
application to astrophysics, INP-ENSEEEIHT/IRIT,
http://rouet.perso.enseeiht.fr/report.pdf

\bibitem[Roques et al.(2003)]{Roques03}
Roques, J.~P., Schanne, S., Von Kienlin, A., et al., \ 2003, \aap, 411, L91


\bibitem[Scargle(1998)]{Scargle98}
Scargle, D. \ 1998, \apj, 504, 405

\bibitem[Smith(2004)]{Smith04}        
Smith, D.~M. \  2004, ESA-SP-552, 45

\bibitem[Strong \& Moskalenko(1998)]{Strong98}                
 Strong, A.~W. \& Moskalenko, I.~V., \ 1998, \apj, 509, 212

\bibitem[Strong et al.(1999)]{Strong99}        
Strong, A.~W., Bloemen, H., Diehl, R., et al., \ 1999, Astrophys. Lett. Commun., 39, 209

\bibitem[Strong et al.(2000)]{Strong00}        
Strong, A.~W., Moskalenko, I.~V., \&  Reimer, O. \ 2000, \apj, 537, 763

\bibitem[Strong et al.(2005)]{Strong05}        
Strong, A.~W., Diehl, R.,  Halloin, H., et al., \ 2005, \aap, 444, 495

\bibitem[Strong et al.(2004)]{Strong04}        
Strong, A.~W., Moskalenko, I.~V., \&  Reimer, O. \ 2004, \apj, 613, 962

\bibitem[Strong et al.(2007)]{Strong07}        
Strong, A.~W., Moskalenko, I.~V., \& Ptuskin, V.~ S., \ 2007, Ann. Rev. Nucl. Part. Sci., 57, 285

\bibitem[Strong(2010)]{Strong2010}
 Strong, A.W.,  Proceedings of the ICATPP Conference on Cosmic Rays for Particle and Astroparticle Physics, 2010, to be published by World Scientific (Singapore), arXiv:1101.1381.

\bibitem[Strong et al.(2010)]{Strong10}        
Strong, A.~W., Porter, T~A., Digel, S.~W., J\`ohannesson, G., Martin, P., Moskalenko, I.~V., Murphy, E.~J., \& Orlando, E.
\ 2010, \apjl, 722, L58

\bibitem[Sturner et al.(2003)]{Sturner03}
Sturner S.~J., Schrader C.~R., Weidenspointner, G., et al., \ 2003, \aap, 411, L81
 
\bibitem[Su, Slatyer \& Finkbeiner(2010)]{Su10}     
Su, M., Slatyer, T.~R., \&Finkbeiner, D.~P. \ 2010, \apj, 724, 1044

\bibitem[Terrier et al.(2010)]{Terrier10}
Terrier R., Ponti G., B\'elanger G., et al., \ 2010, \apj, 719,1 43


\bibitem[Trotta et al.(2010)]{Trotta10}        
Trotta R.~L., J\'ohannesson, G., Moskalenko I.~V., et al., \ 2010, \apj, 729, 106

\bibitem[Tur, Heger \& Austin(2010)]{Tur10}        
Tur C., Heger,A. \& Austin, S. M. \ 2010, \apj, 718, 357

\bibitem[T\"urler et al.(2010)]{Turler10}        
T\"urler, M., Chernyakova, M.,  Courvoisier, T.~J.-L., et al., \ 2010, \aap, 512, 49

\bibitem[Tzvetomila(2009)]{Tzvetomila09}
Tzvetomila, S. \ 2009, Parallel triangular solution in the out-of-core 
multifrontal approach  for solving large sparse linear systems, http://pantar.cerfacs.fr/6-26642-PhD-Dissertations.php

\bibitem[Valinia et al.(2000)]{Valinia00}
Valinia, A., Kinzer, R.~L., \& Marshall, F.~E. \ 2000, \apj, 534, 277
        
\bibitem[Vedrenne et al.(2003)]{Vedrenne03}
Vedrenne, G., Roques, J.~P., Schonfelder, V., et al., \ 2003, \aap, 411, L63

\bibitem[Yamasaki et al.(1997)]{Yamasaki97}        
Yamasaki, N.~Y., Ohashi, T., Takahara, F., et al., \ 1997, \apj, 481, 821

\bibitem[Wang et al.(2007)]{Wang07}        
Wang, W., Harris, M.~J., Diehl, R., et al., \ 2007, \aap,  469, 1005

\bibitem[Wang et al.(2009)]{Wang09}        
Wang, W., Lang, M.~G., Diehl, R., et al., \ 2009, \aap, 496, 713

\bibitem[Weidenspointner et al.(2008)]{Weidenspointner08}
Weidenspointner, G., Skinner, G., Jean, P., et al., \ 2008, \nat, 451, 159

\bibitem[Woosley \& Heger.(2007)]{Woosley07}        
Woosley, S.~E., \& Heger, A., \ 2007, \physrep,  442, 269


\end{thebibliography}
\end{document}